\documentclass[a4paper,fleqn,usenatbib]{mnras}
\usepackage{threeparttable}
\makeatletter
\newlength{\abovecaptionskip}%
\makeatother
\usepackage{threeparttable}%

\usepackage[T1]{fontenc}
\usepackage{ae,aecompl}

\usepackage{graphicx,soul}	
\usepackage{amsmath}	
\usepackage{amssymb}	

\title[the stellar halo]{Mapping the Milky Way with LAMOST II: the stellar halo}

\author[Y. Xu et al.]{
Yan Xu$^{1}$\thanks{E-mail:xuyan@bao.ac.cn},
Chao Liu$^{1}$\thanks{E-mail:liuchao@bao.ac.cn},
Xiang-Xiang Xue$^{1}$,
Heidi Jo Newberg$^{2}$,
Jeffrey, L. Carlin$^{3}$,
\newauthor Qi-Ran Xia$^{1}$ ,
Li-Cai Deng$^{1}$,
Jing Li$^{4}$,
Yong Zhang$^{5}$,
Yonghui Hou$^{5}$,
Yuefei Wang$^{5}$,
\newauthor Zihuang Cao$^{1}$
\\
$^{1}$Key Laboratory of Optical Astronomy, National Astronomical Observatories, Chinese Academy of Sciences,Da tun Road 20A,\\ Beijing 100012, China\\
$^{2}$Department of Physics, Applied Physics and Astronomy, Rensselaer Polytechnic Institute, Tory, NY 12180, USA\\
$^{3}$LSST, 950 North Cherry Avenue, Tucson, AZ 85719, USA\\
$^{4}$ Key Laboratory of Galaxies and Cosmology, Shanghai Astronomical Observatory, Chinese Academy of Sciences, Shanghai 200030, China\\
$^{5}$ Nanjing Institute of Astronomical Optics $\&$ Technology, National Astronomical Observatories,\\ Chinese Academy of Sciences, Nanjing 210042, China\\
}

\date{Accepted XXX. Received YYY; in original form ZZZ}

\pubyear{2017}

\begin{document}
\label{firstpage}
\pagerange{\pageref{firstpage}--\pageref{lastpage}}
\maketitle

\begin{abstract}

The radial number density and flattening of the Milky Way's stellar halo is measured with $\mathrm{5351}$ metal-poor ([Fe/H]$<-1$) K giants from LAMOST DR3, using a nonparametric method which is model independent and largely avoids the influence of halo substucture. The number density profile is well described by a single power law with index $5.03^{+0.64}_{-0.64}$, and flattening that varies with radius. The stellar halo traced by LAMOST K giants is more flattened at smaller radii, and becomes nearly spherical at larger radii. The flattening, $q$, is about 0.64, 0.8, 0.96 at $r=15$, 20 and 30 kpc (where $r=\sqrt{R^2+\left[Z/q\left(r\right)\right]^2}$), respectively. Moreover, the leading arm of the Sagittarius dwarf galaxy tidal stream in the north, and the trailing arm in the south, are significant in the residual map of density distribution. In addition, an unknown overdensity is identified in the residual map at (R,Z)=(30,15) kpc.

\end{abstract}




\section{Introduction}
The spatial distribution of the halo
tracers, specifically the radial number density profile and the flattening of the tracer population, is crucial for understanding the formation history and dark matter distribution in the Milky Way. However, there is little consensus on the radial density and flattening profile of the stellar halo.

In the early studies, based on a few hundred objects (i.e. globular clusters, RR Lyrae variables, blue horizontal branch stars, and K dwarfs, etc.), radial number density profile of the Milky Way halo was usually described as a single power law (SPL) with index of $2.5-3.5$ and the flattening either kept constant $q=\mathrm{0.5-1}$ \citep{Harris1976,Hawkins1984,Sommer-Larsen1987,Preston1991,Soubiran1993,Wetterer1996,Reid1998,Gould1998,Vivas2006} or varied with radius \citet{Preston1991}. However, other studies explored a deviation from the single power law. For instance, \citet{Saha1985} found the density profile traced by RR Lyrae stars followed a broken power law (BPL) with an index of 3 within 25 kpc and an index of 5 beyond 25 kpc.

\begin{table*}\tiny
\begin{threeparttable}
\centering 
\caption{Incomplete list of recent stellar halo profile fits}
\label{previous-works}
\begin{tabular}{lllllll}\hline
  reference &  origin  & tracer & sample size & distance(kpc) & model & parameters                    \\
  \hline\noalign{\smallskip}
 \citet{Iorio2017} & GAIA+2MASS & RR Lyrae & 21600 & $R<28$ & Triaxial & n=2.96, p=1.27, $q=f(r)$, $q_{0}=0.57$, $q_{inf}=0.84$,\\
                   &           &          &       &        &          &                $r_{0}=12.2 kpc$\\
\\
 \citet{Das2016}  &  SEGUE2 & BHB &   & $r_{GC}<70$ & BPL & $n_{in}=3.61$, $n_{out}=4.75$, $r_{break}=29.87$, q=0.72\\
         &        &          &      &                & SPL      &  n=4.65, $q=f(r_{GC})$, $q_0=0.39$, $q_{inf}$=0.81, $r_0=7.32Kpc$\\
\\
  X15 & SEGUE2 & K giants & 1757 & $10<r_{GC}<80$ & BPL      & $n_{in}=2.8$, $n_{out}=4.3$, $r_{break}=29$, q=0.77\\
         &        &          &      &                & Einasto  & n=2.3, $r_{eff}=18$, q=0.77\\
         &        &          &      &                & SPL      &  n=4.4, $q=f(r_{GC})$, $q_0=0.3$, $q_{inf}$=0.9, $r_0=9Kpc$\\
\\
  \citet{Pila-Diez2015} & CFHTS \& INT & near MSTO&  & $r_{GC}<60$ & SPL  &  n=4.3, q=0.79\\
                                &               &         &  &             & Triaxial & n=4.28, q=0.77, $\omega=0.87$\\
                                &              &          &  &             &  BPL     & $n_{in}=2.4$, $n_{out}=4.8$, $r_{break}=19$, q=0.77\\                                        &              &          &  &             &  BPL$_q$     & $n_{in}=3.3$, $n_{out}=4.9$, $q_{in}=0.7$, $q_{out}=0.88$\\
\\
  \citet{Deason2011} & SDSS DR8& BS,BHB     &$\sim20000$ & $4<D<40$  & BPL      & $n_{in}=2.3$, $n_{out}=4.6$, $r_{break}=27$, q=0.6\\
\\
  \citet{Deason2014} & SDSS DR9& BS,BHB     &  & $10<D_{BS}<75$ & BPL      & $n_{outer}=6-10$, $r_{break}=50$\\
                             &         &            &  & $40<D_{BHB}<100$ &          &                                 \\
\\
  \citet{Watkins2009}              & Stripe82 & RRly & 417 & $5<r_{GC}<117$ & BPL & $n_{in}=2.4$, $n_{out}=4.5$, $r_{break}=25$\\
\\
  \citet{Sesar2011} & CFHTLS & near MSTO &  27544 & $D<35$ & BPL & $n_{in}=2.62$, $n_{out}=3.8$, $r_{break}=28$, q=0.7\\
\\
  \citet{Juric2008} & SDSS & MS &  & $D<20$ & SPL & $n=-2.8$, q=0.64\\
\\
  \citet{Bell2008}  & SDSS & MS & 4 million & $D<40$ & SPL & $2<n<4$, $0.5<q<0.8$\\
\\
  \citet{Siegel2002} & Kapteyn &  & 70000 &           & SPL & $n=2.75, q=0.6$\\
\\
  \citet{Robin2000} &  PB  &    &           &        & SPL & $n=2.44, q=0.76$\\
\hline 
\end{tabular}
\begin{tablenotes}
\item[-] CFHTLS (Canada-France-Hawaii Telescope Legacy Survey) 
\item[-] near MSTO (near main sequence turnoff stars)
\item[-] INT (Isaac Newton Telescope)
\item[-] PB (Pencil beams from different observations)
\item[-] Kapteyn (seven Kapteyn selected areas)
\end{tablenotes}
\end{threeparttable}
\end{table*}

With the advent of large sky suveys, the halo density profile can be studied using an expanded sample with larger sky coverage and out to larger distances. Measurements from recent literature are summarized in Table~\ref{previous-works}. These studies show that more complicated, multi-parameter models of the global number density profile are often required to fit the expanded datasets. Some researchers found that the density profile follows a single power law with a constant flattening \citep{Robin2000, Siegel2002,Bell2008,Juric2008}, while other studies claimed that a broken power law or equivalent Einasto profile \citep{Einasto1965} with a constant flattening is more suitable to describe the observed data \citep{Sesar2011,Deason2011}. \citet{Deason2014} found that there are two breaks located at 27 kpc and 50 kpc, respectively, in the density profile. \citet[herafter X15]{Xue2015}
clarified that a broken power law or Einasto profile with constant flattening can fit the density profile of SEGUE K giants or, alternatively, a single power law with variable flattening can also fit the data well . They attributed the apparent radial change in power law index to the variation of flattening. \citet{Pila-Diez2015} also found the halo is more flattened within 19 kpc, and becomes less flattened beyond 19 kpc.

The results of previous works listed in Table~\ref{previous-works} come with some caveats. First, some tracers (such as RR Lyrae stars) are quite rare objects in the Galactic halo, with total sample size of only a few hundreds. Second, different tracers may trace different stellar populations. For example, since  on the horizontal branch is correlated with metallicity, BHBs and RR Lyraes represent different stellar populations. Finally, for all populations, the density profile is not well measured at large distances, preventing tight constraints of the shape of the outer halo.

Furthermore, the power law index and flattening parameters in the density models are degenerate to some degree. X15 found that the SDSS K giants can be empirically characterized using several models with constant or variable flattening, but it is impossible to distinguish which one is correct due to degeneracy between the models. Therefore, it is crucial to determine power law index and flattening independently.

An important consideration in determining the density profile of the stellar halo is that a large fraction of halo stars are associated with substructure. \citet{Watkins2009} indicated that 60\% of the stars in their sample belong to substructure. \citet{Sesar2010} found that 20\% of the stars within 30 kpc belongs to substructure, and beyond that distance the fraction is larger. X15 found that the substructure does affect the halo density profile. However, it is difficult to identify each individual member star of a given substructure. Fitting the density profile of the smooth component alone is often accomplished by cutting out regions of  the sky that contain significant density substructure to avoid their influence. The problem of this approach is that
most significant substructures (such as the Sgr stream and Virgo overdensity) are very extended. As a consequence, if we exclude the Sgr stream and Virgo overdensity according to the \citet{Bell2008} criteria, most stars higher than $b=60^\circ$ are masked. This makes it difficult to constrain $q$, since $q$ is sensitive to density comparisons between the low and high Galactic latitudes.

Liu et al. (2017 hereafter Paper I) developed a statistical method to derive the stellar density profile from the spectroscopic survey data. They qualitatively found that the metal-poor ([Fe/H]$<$-1) halo-like RGB stars have a quite oblate shape within a Galatocentric radius of about 20 kpc. When looking at radii larger than 30 kpc, the shape of the halo density profile becomes nearly spherical. This is evidence that the flattening of the stellar halo varies with radius, rather than following a broken power law.

In this work, we quantitatively investigate this with more carefully selected K giant samples and a well defined approach. A nonparametric technique is adopted to measure the density profile of the Galactic halo with The Large sky Area Multi-Object fiber Spectroscopic Telescope (LAMOST) Survey metal poor ([Fe/H]$<-1$) K giants as far as 35 kpc. Using this method, we fit the iso-density surface of LAMOST K giants directly without presuming functional forms that describe the density and variation of flattening. In addition, the influence of the outliers caused by substructure will be eliminated by fitting the median value of the iso-density surface if the substructure do not dominate in a particular bin. So the advantage of the nonparameteric method is that it can estimate the power law index and the flattening independently, while avoiding the influence of substructures.

In next section, we describe the K giant sample and the stellar density estimation. The nonparametric technique and the halo profile are presented in section 3. Finally, in section 4 we discuss the substructures that were not included in our fit, and draw brief conclusions.

\section{LAMOST K giants and the stellar density}

\subsection{LAMOST K Giants}
LAMOST is a 4-meter, quasi-meridian, reflecting Schmidt telescope which was designed for a large spectroscopic survey\citep{Cui2012, Zhao2012, Deng2012}. The stellar parameters are estimated in the LAMOST pipeline (Wu et al. 2011, 2014).

\begin{figure*}
    \includegraphics[width=\columnwidth]{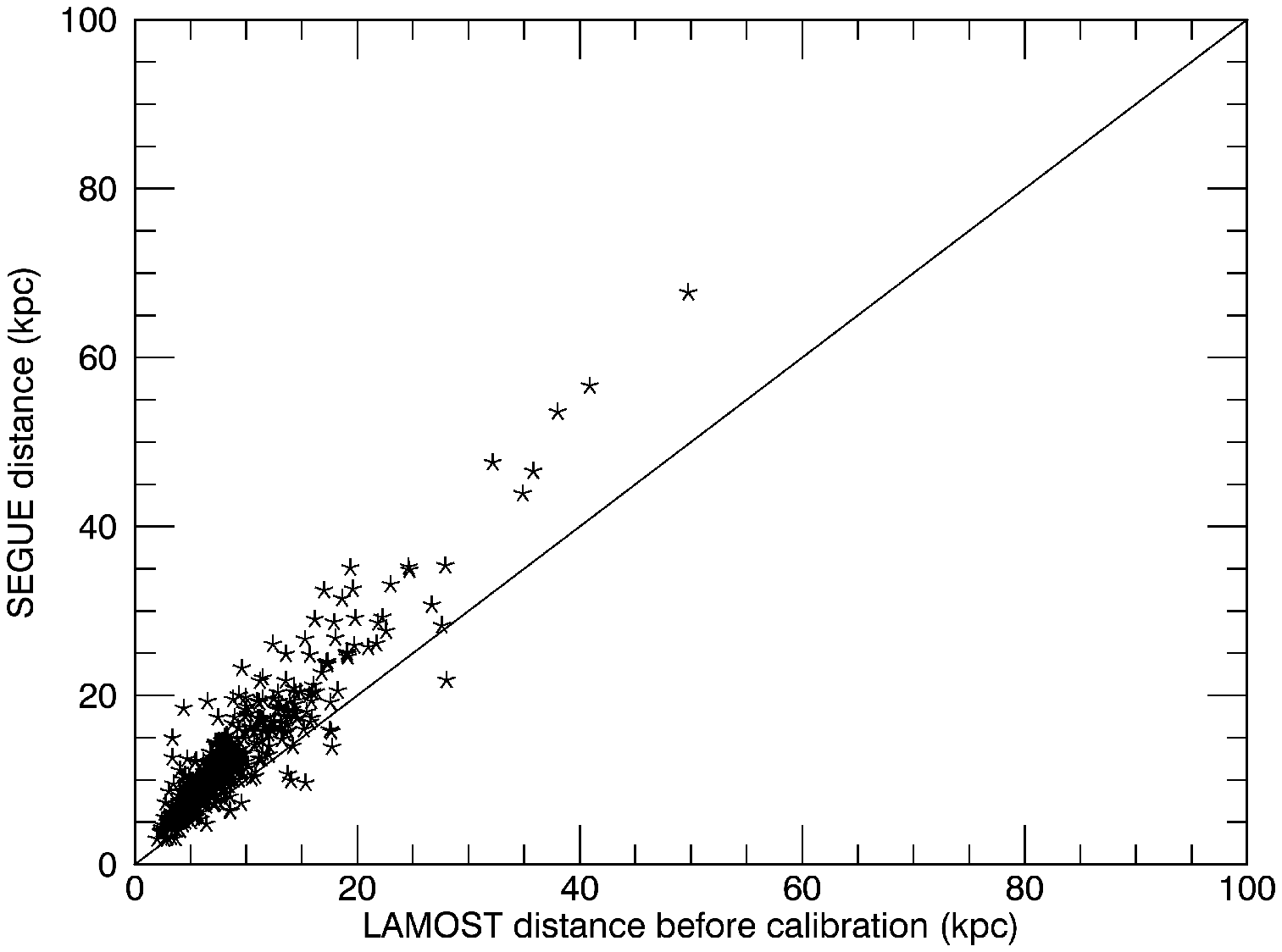}
   \includegraphics[width=\columnwidth]{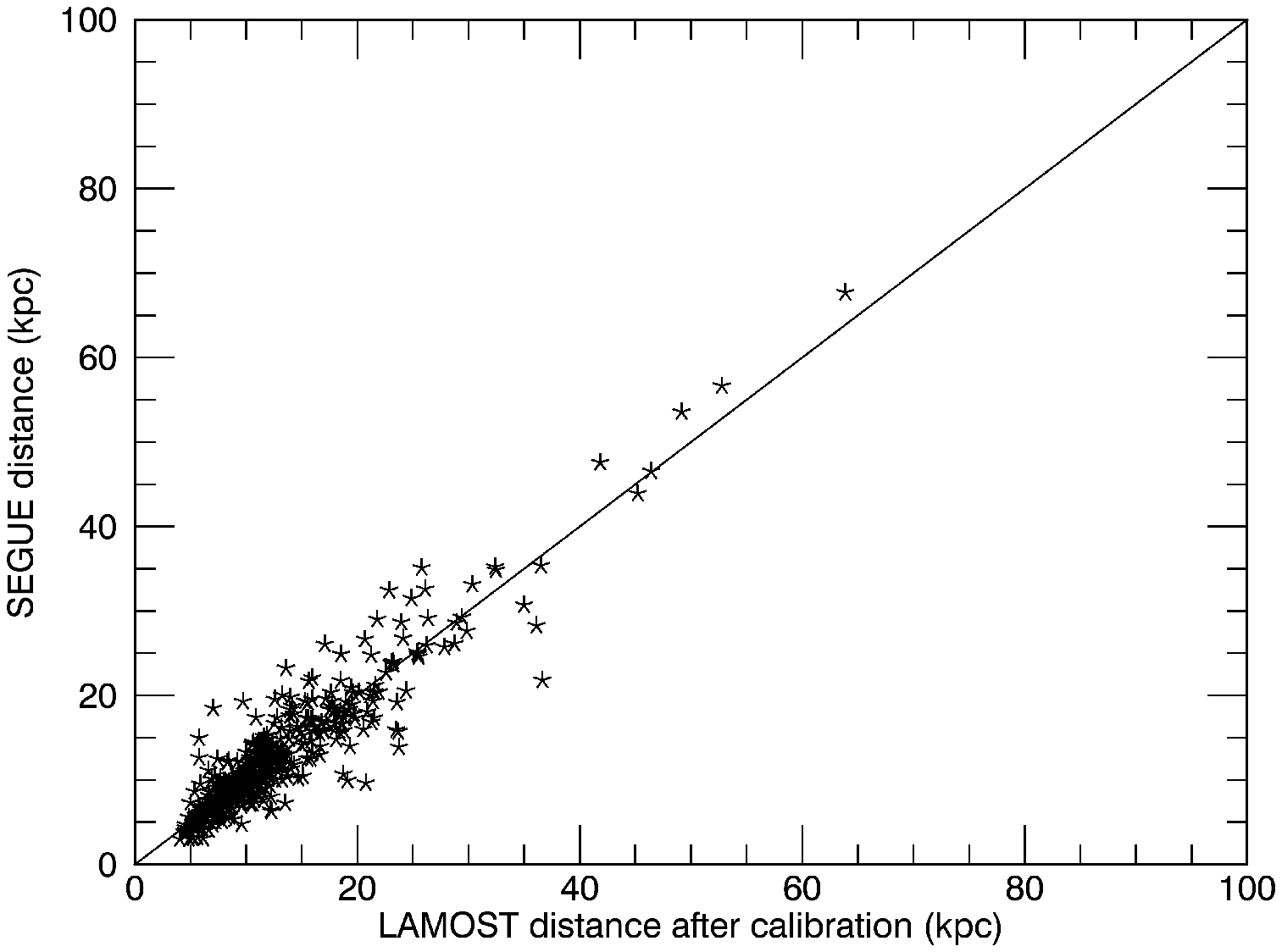}
    \caption{The distances of 400 common K giants in LAMOST and SEGUE. The left panel shows there is an obvious bias in distance caused by the photometric difference between PS1 and SDSS. The right panel shows the distances derived from LAMOST spectra are consistent with those from SEGUE spectra after calibration.}
    \label{calibration}
\end{figure*}

The LAMOST Data Release 3 (DR3) catalogue contains over 4 million stellar spectra, of which 350,000 stars are identified as K giants \citep{Liu2014}. Contamination from red clump stars is removed using the technique for identifying red clump stars provided by Wan et al. (2015), Tian et al. (2016), and Wan et al. (2017). The distances of these K giants are derived based on the fiducial isochrones in SDSS extinction-corrected $g-r$ $\it vs.$ absolute r band magnitude, $M_r$, Pan-STARRS1 $g_{p1}$ and $r_{p1}$, and LAMOST [Fe/H], using Bayesian method described in \citet{Xue2014}. The photometry for the K giants is corrected for extinction based on $E(B-V)$ of \citet{Schlegel1998}, and the coeficients in Table 6 of Schlafly et al. (2011).  K giants with $E(B-V)>0.25$ are eliminated to avoid the influence of poor extinction correction in areas of high extinction; the typical $g-r$  error caused by the extinction correction is about 0.017 using this criterion (Xue et al. 2014). The errors in colour, magnitude, and metallicity are propagated into the distance uncertainty. The systematic distance bias which is caused by the difference between SDSS $g$ and $r$ and Pan-STARRS1 $g_{p1}$ and $r_{p1}$ is calibrated using $\sim$ 400 common K giants, as shown in Figure~\ref{calibration}. Including all of these factors, the distance estimates for LAMOST K giants have a typical uncertainty of about 15\% .

\begin{figure}
    \includegraphics[width=\columnwidth]{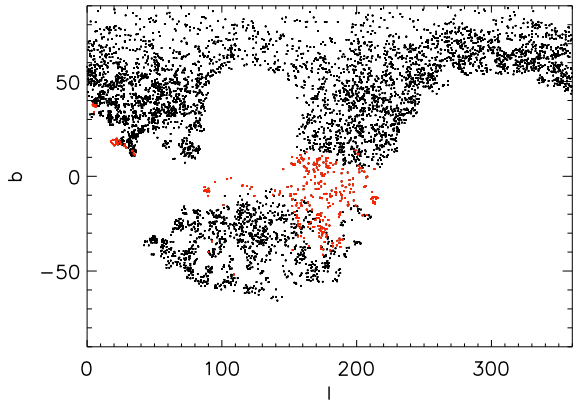}
    \caption{Sky coverage of the selected LAMOST K giants([Fe/H]$<-1,M_{r,p1}<-0.5$,$K<14^m.3$). The red points label K giants with $E(B-V)>0.25$ which are eliminated from our sample.}
    \label{skycoverage}
\end{figure}

\begin{figure}
        \includegraphics[width=\columnwidth]{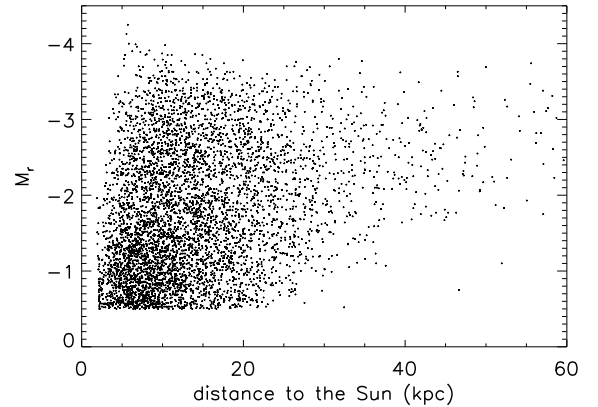}
    \caption{Distribution of $M_{r,p1}$ vs. distance for the selected K giants([Fe/H]$<-1,M_{r,p1}<-0.5$,$K<14^m.3$,$E(B-V)>0.25$)}
    \label{M_K_dist}
\end{figure}

To eliminate contamination from disk stars, we cull K giants with [Fe/H]$<-1$. Because the spectroscopic selection effects are corrected based on the photometry from The Two Micron All-Sky Survey (2MASS, Cutri et al. 2003), we impose a faint magnitude limit of $K<14^m.3$. In addition, the absolute $r_{p1}$ band magnitude, $M_{r,p1}<-0.5$, is restricted in order to maintain the completeness of $M_{r,p1}$ up to distances of 35 kpc. We also constrain $E(B-V)<0.25$ in our K giant samples, to avoid unreliable reddening at low latitudes. After all of these cuts, there are 5351 K giants with [Fe/H]$<-1$, $M_{r,p1}<-0.5$, $K<14^m.3$, and $E(B-V)<0.25$ in our sample. Figure~\ref{skycoverage} shows the sky coverage of the selected K giant sample. The red points are eliminated from our sample because their $E(B-V)$ is larger than 0.25, which excludes the stars in the range of $-10^\circ<b<10^\circ$. But our sample still has a sizeable sky coverage in the range $b=-60^\circ \sim 90^\circ$, which is sufficient for constraining the global structure of the Galactic stellar halo. Figure~\ref{M_K_dist} illustrates the distribution of the K giants in the $M_r$ vs. distance plane. The total sample, with $M_{r,p1}$ in the range $[-4^m, -0.5^m]$ is barely complete up to a distance of 35 kpc (The completeness here is defined as a volume within which the coverage of the absolute magnitude of the samples should be similar at any place.). Beyond 35 kpc, we begin to lose intrinsically fainter stars. In addition, the number of available stars declines dramatically. Therefore, in order to keep a relatively large samples in a relatively large distance, we select a lower limit of $M_{r,p1}=-0.5^m$. A lower limit fainter than this would mean that we may not have complete coverage out to 35 kpc, which is critical for investigating the breaks of power law index.

\subsection{Estimation of the stellar density}
Paper I described a statistical method to derive the stellar density profile along each line-of-sight using spectroscopic survey data. Selection effects due to target selection and observational biases are taken into account in the density profile estimation. Here we briefly describe the method and suggest that the readers to refer to Paper I for more details about the approach.

We first assume that the photometric survey data is complete for the stars brighter than its limiting magnitude. Then we assume that, in a given line-of-sight which is defined by the Galactic coordinates ($l$, $b$) with solid angle $\Omega$, the probability of finding a star at distance $D$ in the spectroscopic survey data is approximately same as it is in the photometric data in a small region in the --magnitude plane, i.e.,
\begin{equation}\label{eq:pp}
p_{ph}(D|c,m,l,b)=p_{sp}(D|c,m,l,b),
\end{equation}
where $c$ and $m$ stand for the  index and apparent magnitude, respectively.

The two probability density functions can be expressed in terms of the stellar densities derived from the photometric data and the spectroscopic data, respectively.  Note that the photometric data is presumed to be complete, while the spectroscopic data is not.  Then, based on Eq.~(\ref{eq:pp}) the two densities, denoted as $\nu_{ph}$ and $\nu_{sp}$ respectively, are associated with a factor $S$, which represents the selection function of the spectroscopic data:
\begin{equation}\label{eq:nuPDF}
\nu_{ph}(D|c,m,l,b)=\nu_{sp}(D|c,m,l,b)S^{-1}(c,m,l,b),
\end{equation}

Integrating over $c$ and $m$, we obtain the stellar density profile along the given line-of-sight:
\begin{equation}\label{eq:nuall}
\nu_{ph}(D|l,b)=\iint{\nu_{sp}(D|c,m,l,b)S^{-1}(c,m,l,b)dcdm}.
\end{equation}
Paper I proved that Eq.~(\ref{eq:nuall}) is held not only for the whole dataset, but also for a selected stellar population.

The selection function $S(c,m,l,b)$ can be derived from
\begin{equation}\label{eq:S2}
S(c,m,l,b)={n_{sp}(c,m,l,b)\over{n_{ph}(c,m,l,b)}},
\end{equation}
where $n_{sp}(c,m,l,b)$ and $n_{ph}(c,m,l,b)$ are the star counts of the spectroscopic and photometric data in the -magnitude plane, respectively.

To obtain $\nu_{ph}$ we need only estimate $\nu_{sp}$ from the observed spectroscopic data according to Eq.~(\ref{eq:nuall}). We apply a kernel density estimation (KDE) to estimate $\nu_{sp}$ along a line-of-sight considering that 1) the number of spectroscopic stars is usually limited along a line-of-sight and 2) there is uncertainty of the distance. Then we can write
\begin{equation}\label{eq:KDE2}
\nu_{sp}(D|c,m,l,b)={1\over{\Omega D^2}}\sum_{i}^{n_{sp}(c,m,l,b)}{p_i(D)},
\end{equation}
where $p_i(D)$ is the probability density function of $R$ for the $i$th star.
Note that in this form the derived $\nu_{ph}$ is a continuous function of $D$.

For a large survey like LAMOST, thousands of ``plates" (used here to mean an observation with in this case 4000 fibers along a line-of-sight, even though the fiber positioners do not require production of a physical metal plate as in other surveys) have been observed. In principle, we can derive $\nu_{ph}$ for all plates and we can obtain $\nu_{ph}(D)$ at any arbitrarily selected distance $D$. However, at the position where no spectroscopic stars are sampled, the derived stellar density may suffer from large uncertainty. Hence, in practice, we only use the stellar density at the position in which a spectroscopic star is located, so that the precision of stellar density is preserved.

The advantage of this method is that no specific analytical form of the stellar density profile of the Milky Way is required. In other words, the stellar density estimation does not depend on the selection of a stellar density model of the Galaxy. For this reason, in this work, we are able to map the stellar halo density profile in a non-parametric manner.

Paper I did performance assessments and found that the typical error in the derived stellar density is less than $\sigma_\nu/\nu=0.4$. When the samples are very sparse along a line-of-sight, the derived stellar density profile may have some systematic bias. However, this systematic bias can be blurred when we solve for the Galactic structure by combining multiple lines-of-sight. But even in the worst case, when the systematic bias reaches $\sigma_\nu/\nu\sim0.5$--$1$, it can only produce deviations of about 10\% in the structural parameter estimates, according to the simulations presented in Paper I. Therefore, we infer that the accuracy of the stellar density estimates based on Paper I should be sufficient for our the study of the stellar halo.

We then estimate the stellar density at the positions of the selected K giant samples with [Fe/H]$<-1$, $M_{r,p1} < -0.5$, $K < 14^m.3$, and $E(B-V)<0.25$. Because these samples are complete out to 35 kpc, the stellar density estimates should be reliable within this volume.

\subsection{Number Density Map}
\begin{figure}
    \includegraphics[width=\columnwidth]{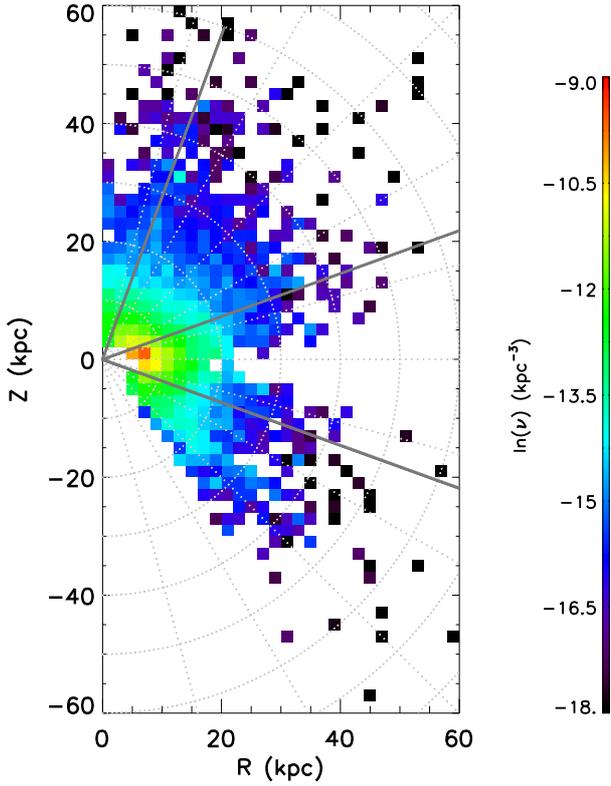}
    \caption{The observed number density distribution of the selected K giants on a natural logarithmic scale in a Galactocentric cylindrical coordinate system.
The pixel size of the map is $2\times2$ (kpc). The number density in each pixel represents the median of the number densities of K giants in this bin. The number densities are  coded according to the colour bar; the values of $\ln(\nu)$ outside the range of (-9, -18) saturate to black and red, respectively. White pixels represent regions of the Galaxy that were not sampled. The grey points show concentric circles around the Galactic centre with radii from 10 kpc to 60 kpc.
The grey lines indicate two wedge-like subsets at $\theta<20^\circ$ and $\theta>70^\circ$ $[\theta=arctan(|Z|/R)]$, respectively. 
}
    \label{hessRZ}
\end{figure}

\begin{figure}
    \includegraphics[width=\columnwidth]{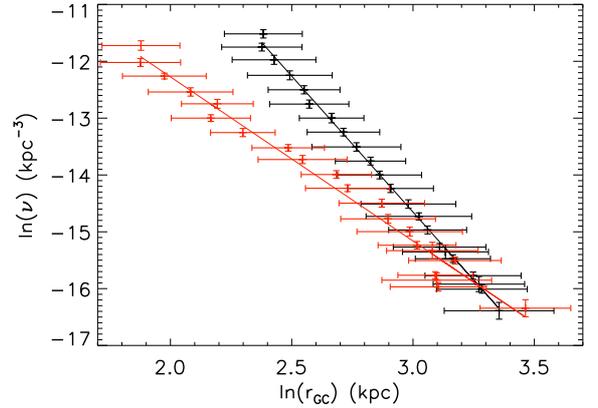}
    \caption{
This plot shows density distributions of the two wedge-like subsets defined in Figure~\ref{hessRZ}. 
The data points indicate the mean values of ($ln(\nu), ln(r_{GC})$)   in each $ln(\nu)$ bin listed
 in Table ~\ref{parameter}. The vertical and horizontal error bars represent the dispersions, defined by standard deviation, in $ln(\nu)$ and $ln(r_{GC})$, respectively. The black and red dots represent the wedges with 
 $\theta<20^\circ$ and $\theta>70^\circ$, respectively. The best fit slope for the data with $\theta<20^\circ$ is -4.78, while it becomes -2.89 for the data with $\theta>70^\circ$.
}                                                                                                             
    \label{densityprofile}
\end{figure}                                                                                                  

\begin{figure}
        \includegraphics[width=\columnwidth]{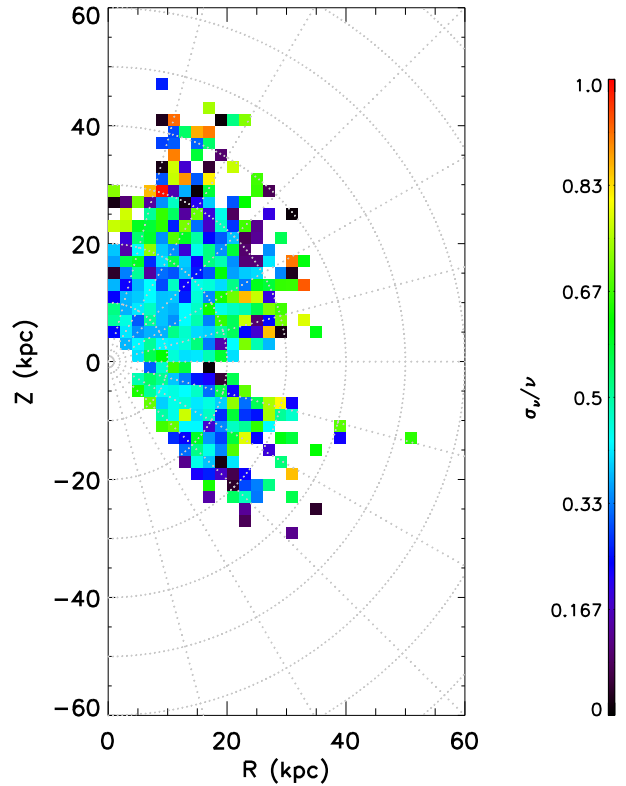}
    \caption{This plot shows the ratio between the dispersion of the number density and the number density in each R-Z bin of Figure~\ref{hessRZ}. The dispersion of the number density is described by MAD (median of absolute deviation) of the number density. The dispersion is calculated only for bins with more than one star; consequently, some bins are missing in this plot comparing with Figure~\ref{hessRZ}.}
    \label{ratioMAD_RZ}
\end{figure}

\begin{figure}
    \includegraphics[width=\columnwidth]{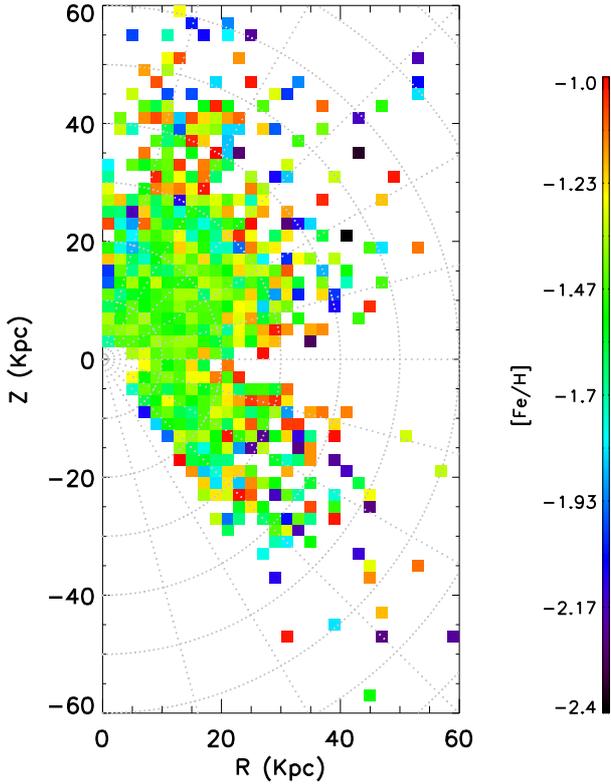}
    \caption{The median [Fe/H] of the selected K giants in each (R,Z) grid.}
    \label{hessRZfeh}
\end{figure}

We explore the stellar number density map as a function of ($R, Z$), where $R$ is radial distance on the Galactic plane ($R=\sqrt{X^2+Y^2}$) and $Z$ is the vertical distance to the Galactic mid-plane. The number density of K giants ($\nu$) at each spatial position is calculated using the method described in Paper I. In Figure~\ref{hessRZ}, a tomographic map is constructed by calculating median for all values of the density of K giants ($median(\nu)$) at each R-Z pixel. 
In this map, there is no obvious exponential disc component near the Galactic plane due to the low metallicity selection. The isodensity contours look more like oblate ellipses rather than spherical within 30 kpc. Although the sample selection, distance estimates and interstellar extinction correction are different, our density map is quite consistent with the results of Paper I (see their figure 14). This indicates that our estimate of the halo density profile is quite robust.

In light of paper I, the stellar halo may be oblate in inner region and spherical in outer region. This can be directly tested by comparing the stellar density profile along $ln(R)$ with that along $ln(Z)$. Now consider that the profile follows a power-law:
\begin{equation}\label{eq:powerlaw}
\ln(\nu)\propto n/2\;ln[R^2+(Z/q)^2]
\end{equation}
When the density profile is measured along $ln(Z)$, it becomes
\begin{equation}\label{eq:powerlaw_lnZ}
\ln(\nu)\propto n\;ln(Z)-n\;ln(q)
\end{equation}
And when the density profile is along $ln(R)$, it reads
\begin{equation}\label{eq:powerlaw_lnR}
\ln(\nu)\propto n\;ln(R)
\end{equation}

If the stellar halo has a constant q, then  it is easy to infer that the stellar density profiles along $ln(R)$ and $ln(Z)$ should be parallel with each other. If q increases with increasing radius, then the slope in equation~\ref{eq:powerlaw_lnZ} would not be n any more and hence the two density profiles should not be parallel any more.

Figure~\ref{densityprofile} shows the result of the test. We select the data located within $\theta>70^\circ (\theta=arctan(|Z|/R))$ in Figure~\ref{hessRZ} to represent the stellar density profile roughly along ln(Z) (displayed as red symbols in Figure~\ref{densityprofile}) and select the data located at $\theta<20^\circ$ to stand for the profile along $ln(R)$ (black symbols in the figure). We fit them with single power-law and obtain that the best fit power index for the subset with $\theta<20^\circ$ is -4.78, while for the subset with $\theta>70^\circ$ is -2.89. It is obvious that the two density profiles are not parallel with each other, meaning that the flattening q must be variable with radius. More quantification of the variation is discusssed in next section.

Figure~\ref{ratioMAD_RZ} shows the relative dispersion of the number density in each (R,Z) pixel. The relative dispersion is defined by (MAD of $\nu$)/median($\nu$). MAD of $\nu$ is median of absolute deviation defined as $median|\nu-median(\nu)|$. 
The relative dispersion is quite uniform in the range $10<r_{GC}<20$ kpc, with a mean relative dispersion of $0.42$. In the region where $r_{GC}>20$ kpc, the relative dispersions have larger scatter, but the mean value is still $0.44$. The dispersion in number density is caused by the intrinsic scatter of the number density due to the non-axisymetric substructures as well as the error induced by correction for selection effects and distance estimation.
From the simulation in paper I, the uncertainty of the stellar density caused by selection effect correction can be as large as $\sigma_\nu/\nu=0.5-1$ in the worst case for a single plate, which will induce similar level random error on the iso-density surface. The relative dispersion in this work is not larger than the estimation in the paper I. This means that the density error caused by selection effect correction is well controlled.

Figure~\ref{hessRZfeh} shows that the distribution of the median [Fe/H] of our K giant samples is quite uniform within 25 kpc, while beyond 25 kpc the distribution of median [Fe/H] is more complicated because of overdensities.

\section{Probing the number density profile using a nonparametric method}
\subsection{Method description}
Previous studies explored the radial number density and flattening profile of the Milky Way's stellar halo by fitting to different parameterized models. X15 shows the number density profile traced by SEGUE K giants can be described using the Einasto profile, a broken power law, or a single power law with constant or variable flattening comparably well, which indicates the degeneracy of the models. Thanks to the large sample and wide sky coverage of LAMOST K giants, it is possible to apply a direct approach to study the structure of the Galaxy with minimal assumptions.  Unlike the popular model-dependent method, we devise a nonparametric method to probe the radial number density and flattening profiles of the Galactic stellar halo. We study the structure of the halo by fitting the outline of each iso-density contour to the density surface, directly.

\begin{figure}
    \includegraphics[width=\columnwidth]{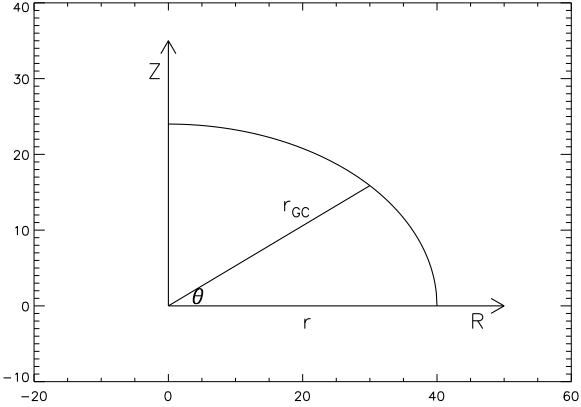}
    \caption{Schematic contour with constant $\ln(\nu)$ in the $R-Z$ plane. It illustrates how we define the quantities in polar coordinates. $r_{GC}$ is Galactocentric distance of an individual star; $\theta=\arctan(Z/R)$ is the angle from the Galactic plane to a given position on the ellipse, as seen from the Galactic centre; and $r$ is the semi-major axis of the ellipse.}
    \label{plotscheme}
\end{figure}

Since many studies show the Galactic halo is oblate, we trace the halo density profile along the semi-major axis ($r$) of an ellipse. Figure~\ref{plotscheme} is a schematic cartoon to illustrate the geometry. $r$ is the semi-major axis of the ellipse defined as $r=\sqrt{R^2+(Z/q)^2}$, where flattening $q$ is the ratio of semi-minor to semi-major axis. $r_{GC}$ is the Galactocentric distance ($r_{GC}=\sqrt{R^2+Z^2}$). If the density profile is characterized by power law, then for each iso-density surface the stellar number density is given by $\nu=\nu_0(1/r)^n$. Therefore, $\ln(\nu)$ has linear relationship with $ln(r)$ i.e. $\ln(\nu)=\ln(\nu_0)-n\;ln(r)$. For a given $\ln(\nu)$, the locus of R-Z pixels also satisfies an elliptical function. 

The semi-major axis, $r$, and flattening, $q$, are determined by fitting the ellipse to each iso-density contour. We find it difficult to constrain $r$ and $q$ in Galacticcentric cylindrical coordinate, because $q$ is very sensitive to the slope of the iso-density surface, and the slope is too steep at low latitude. Therefore, we perform our analysis in Galactocentric polar coordinates:
\begin{equation}\label{eq:ellip_para_func}
\begin{array}{ll}
R=r\cos(\eta),\\
Z=rq\sin(\eta),
\end{array}
\end{equation}
and
\begin{equation}\label{eq:ellip_para_func2}
r_{GC}=\sqrt{R^2+Z^2}.
\end{equation}

Substituting Equation~(\ref{eq:ellip_para_func}) (elliptic parameter equation) into Equation~(\ref{eq:ellip_para_func2}), we find
\begin{equation}\label{eq:ellip_para_func3}
r_{GC}=r\sqrt{1+(q^2-1)sin^2(\eta)}.
\end{equation}

We would like to solve for $r_{GC}$ in terms of the $\theta$ given in Figure~\ref{plotscheme} instead of $\eta$ used in Equation ~(\ref{eq:ellip_para_func}).  Since $\sin(\theta)=Z/r_{GC}$,
\begin{equation}\label{eq:ellip_para_func4}
\sin(\eta)=\frac{r_{GC}\sin(\theta)}{aq}.
\end{equation}
Therefore, we can fit the isodensity contour with Equation~(\ref{eq:ellip_para_func5}):
\begin{equation}\label{eq:ellip_para_func5}
r_{GC}=rq\sqrt{\frac{1}{q^2-(q^2-1)sin^2(\theta)}}.
\end{equation}

\subsection{Fitting to the data}
\begin{figure*}
    \includegraphics[width=12.0cm,angle=90]{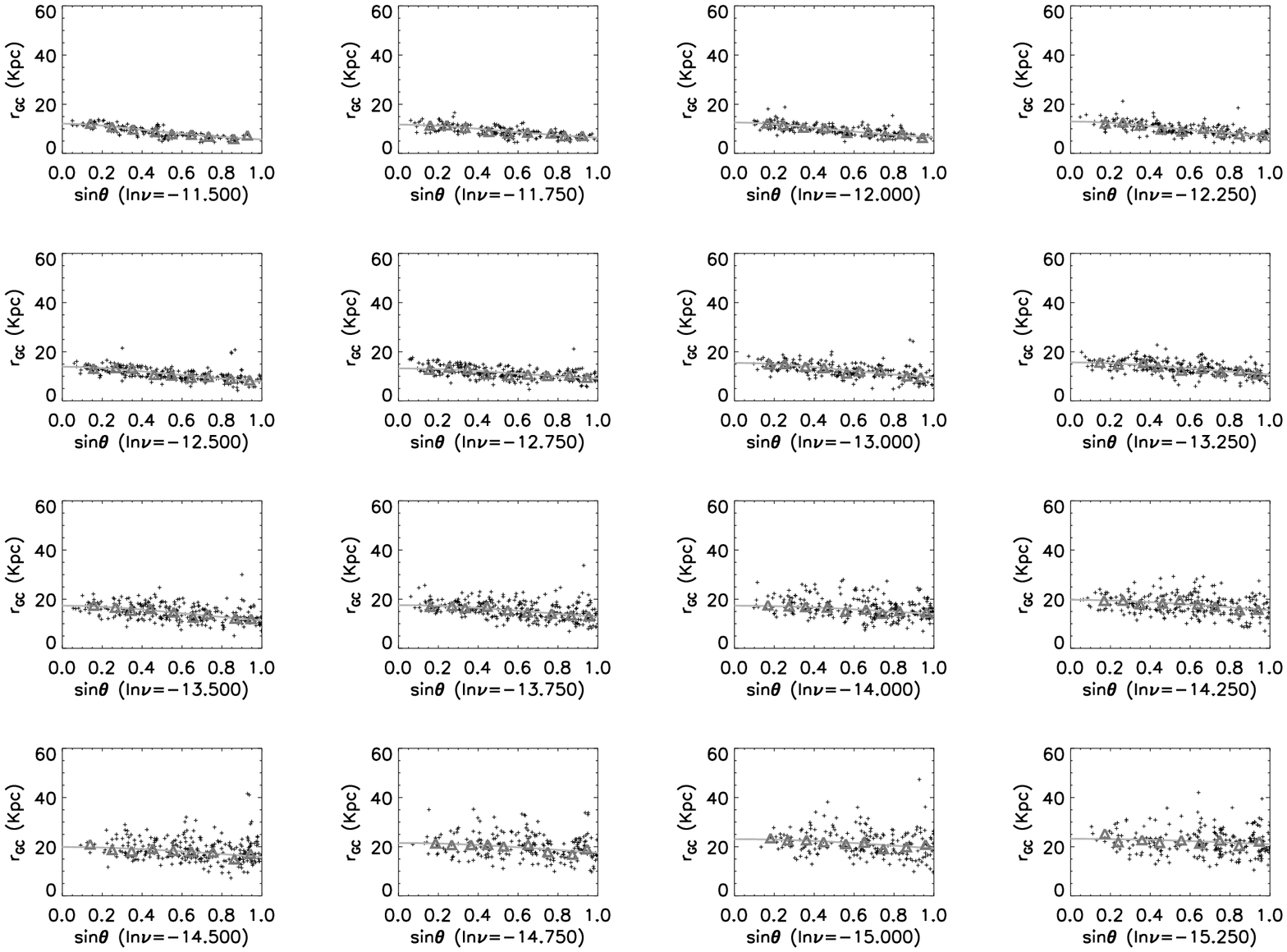}
   \includegraphics[width=12.0cm, angle=90]{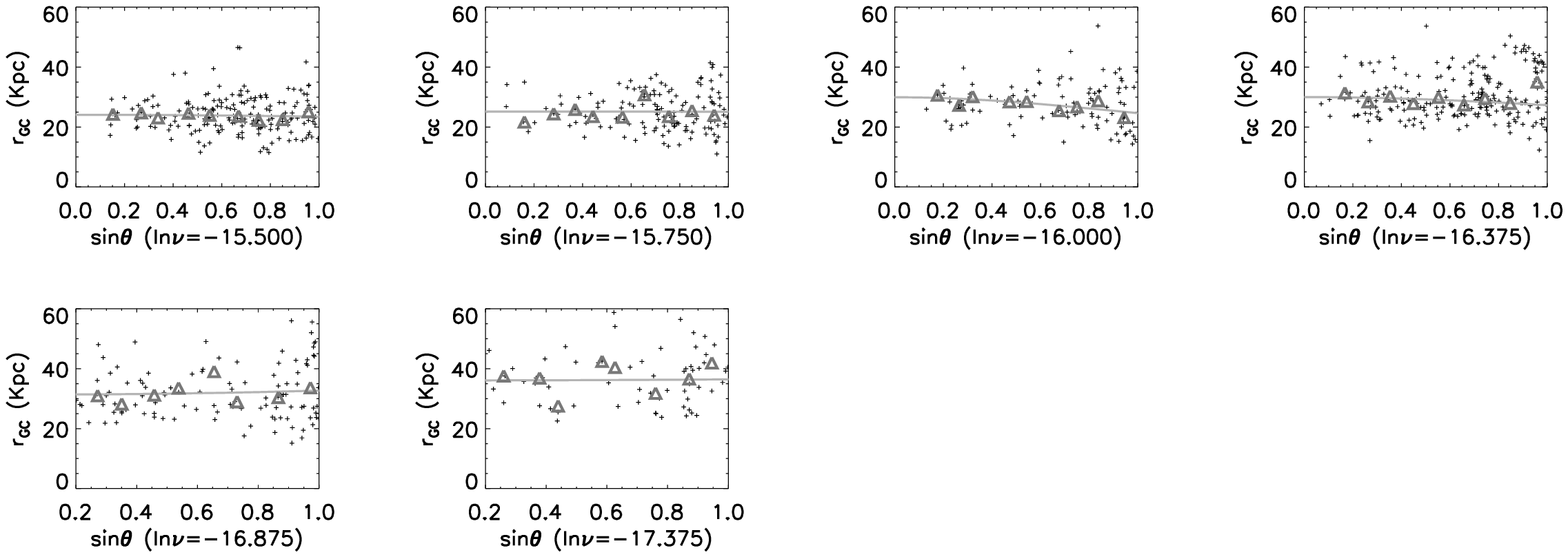}
    \vspace{-5cm}
    \caption{Distribution of K giants (plus signs) in each $\ln(\nu)$ bin in polar coordinates. $r_{GC}$ and $\theta$ are as defined in Figure~\ref{plotscheme}. The mean value of $\ln(\nu)$ in each bin is given in the $X$-axis label for each panel; the width of $\ln(\nu)$ bin is 0.25 except the last three bins, which have a $\ln(\nu)$ bin width of 0.5. In each panel, the median value of $r_{GC}$ in each $\sin\theta$ bin is indicated by a triangle; the $\sin\theta$ bin size is 0.1. The grey curves show the best fit ellipse.}
    \label{rsintheta}
\end{figure*}

Following Paper I, we obtain $\nu$ at the position of each individual K giant.
We divide the K-giant sample into 22 $\ln(\nu)$ bins. The binsize is $0.25$ in the range of $-16.375<\ln(\nu) \le-11.5$, and increases to $0.5$ in the range of $-17.375\le \ln(\nu) \le -16.375$.
The iso-density surfaces and the numbers of stars belonging to them are given in Table~\ref{parameter}.
The distribution of K giants on each iso-density surface and the fitting result is exhibited in Figure~\ref{rsintheta}. The value of each $\ln(\nu)$ bin is given in the X-axis labels.

In general, we should fit the distribution of K giants for each iso-density surface in $(r_{GC} , \sin\theta)$ space.
In fact, we fit the median value of the distribution of the iso-density surface instead of fitting all data points to reduce the infuence of the outliers, many of which could result from halo substructure. The fitting is based on equation (~\ref{eq:ellip_para_func5}), with free parameters $r$ and $q$, using a Markov chain Monte Carlo (MCMC, Foreman-Mackey et al. 2012) method.
The median values have larger scatter in the last few panels, because there are fewer K giants in each $\sin\theta$ bin. Also in the last three panels, the presence of the leading arm Sagittarius in the north strongly influences the last $\sin\theta$ bin; the clump of ourliers aroung $(r_{GC},\sin\theta)=(40,0.95)$, which is primarily composed of Sgr leading tail stars, represents more than half of stars in this bin. So the last bin in the three panels is not used the fitting process. The best-fit parameters $r$ and $q$ are listed in Table~\ref{parameter}.

Note that the bins with $\ln(\nu) > -11.5$ are not used in the analysis. In these bins, the distribution of K giants can be fit by an ellipse or a straight line, because minor axis of the ellipse is small that it is comparable to the scatter of the distribution.

\begin{table}
        \caption{Best-fit parameters $r$, $q$ and their uncertainties for each iso-density contour
        }
        \label{parameter}
        \begin{tabular}{cccccc} 
                \hline
  $\ln(\nu)$ & number &  $r$  & $\sigma_r$ & $q$ & $\sigma_q$                   \\
                \hline
-11.50&165&12.095& 0.969& 0.468& 0.068\\
-11.75&200&11.735& 0.893& 0.546& 0.078\\
-12.00&203&12.595& 1.034& 0.511& 0.074\\
-12.25&221&12.988& 1.083& 0.543& 0.081\\
-12.50&246&13.904& 0.842& 0.565& 0.064\\
-12.75&255&13.203& 0.761& 0.681& 0.077\\
-13.00&244&15.467& 0.750& 0.609& 0.057\\
-13.25&261&15.656& 0.676& 0.683& 0.058\\
-13.50&268&17.375& 0.708& 0.654& 0.051\\
-13.75&298&17.572& 0.641& 0.728& 0.052\\
-14.00&271&17.361& 0.629& 0.803& 0.060\\
-14.25&261&19.748& 0.636& 0.783& 0.051\\
-14.50&267&19.857& 0.610& 0.796& 0.049\\
-14.75&224&21.496& 0.613& 0.814& 0.047\\
-15.00&221&23.059& 0.614& 0.837& 0.045\\
-15.25&216&23.185& 0.578& 0.903& 0.047\\
-15.50&192&24.076& 0.554& 0.974& 0.047\\
-15.75&140&25.141& 0.628& 0.999& 0.050\\
-16.00&106&29.935& 0.602& 0.823& 0.034\\
-16.38&176&29.989& 0.612& 0.910& 0.046\\
-16.88&81&31.349& 0.641& 1.043& 0.052\\
-17.38&52&36.076& 0.677& 1.010& 0.045\\
                \hline
        \end{tabular}
\end{table}

\begin{figure}
   \includegraphics[width=\columnwidth]{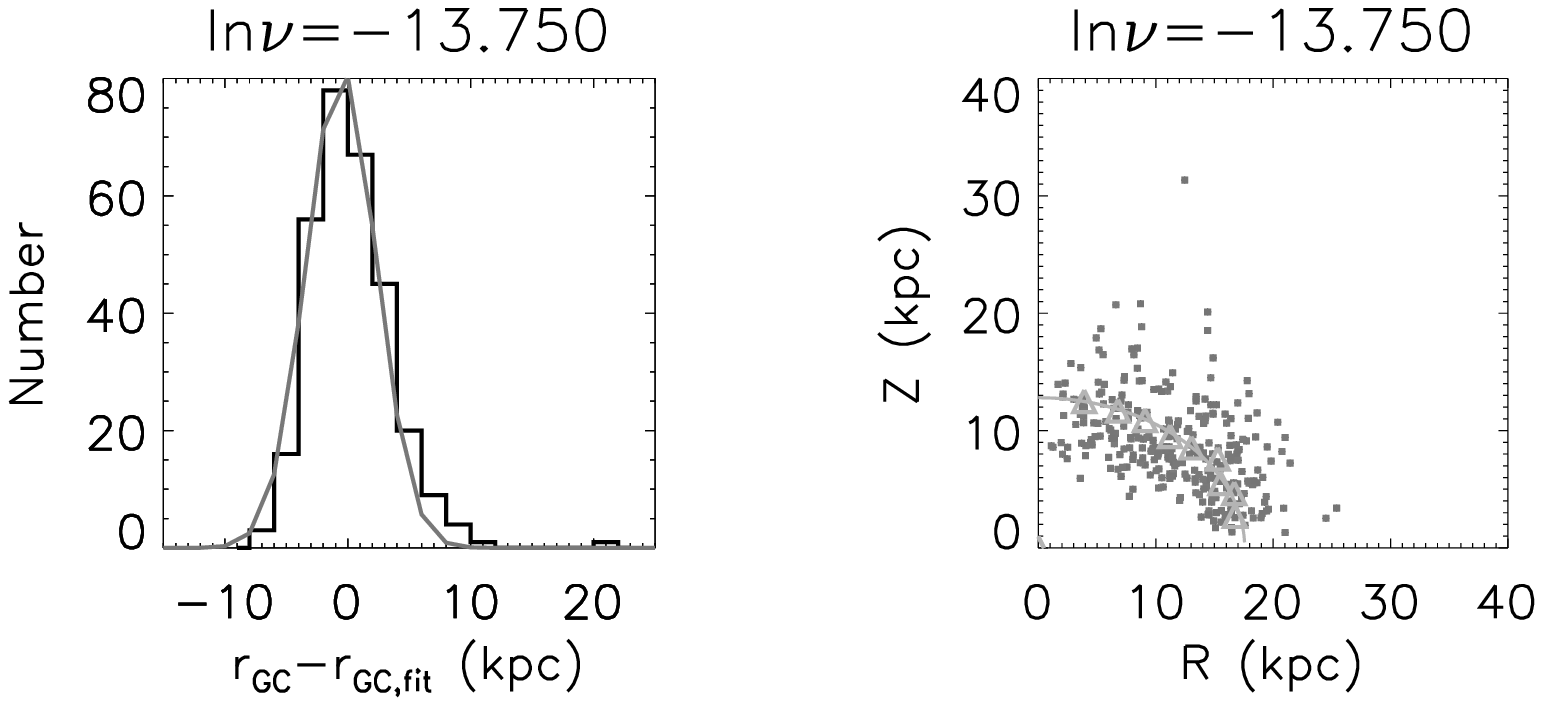}
   \caption{Sample residual fit.  A histogram of fitting residuals ($r_{GC}-r_{GC,fit}$) for the bin with $\ln(\nu)=-13.75$ is shown on the left. The grey curve shows a Gaussian fit to the residual distribution.  
The right panel shows the K giant stars within this iso-density surface and their fitting result in the $R-Z$ plane. The plus signs represent K giants in the bin with $\ln(\nu)=-13.75$. The triangles are the same median values which are shown in Figure~\ref{rsintheta}. The triangles are plotted on this plot according to their polar coordinates($r_{GC},\theta$). The curve shows the outline of the relation: $r=\sqrt{R^2+(Z/q)^2}$.}
   \label{RZ_m13.5}
   \end{figure}

In order to demonstrate the performance of the model fitting, we take a case of $\ln(\nu)=-13.75$ as an example. The left panel of Figure~\ref{RZ_m13.5} shows the deviations from the fit for the data in the $\ln(\nu)=-13.75$ bin ($r_{GC}-r_{GC,fit}$). The distribution of $r_{GC}-r_{GC,fit}$ is well described by Gaussian with $\sigma=2.8$, with the exception that the right wing is slightly over-populated.  This may be due to the presence of density substructures in the halo. Furthermore, we transfer the fitting result back to $R-Z$ space in the right panel of Figure~\ref{RZ_m13.5}. As expected, most K giants in this $\ln(\nu)$ bin are roughly distributed in the shape of an ellipse. The outline of the fitting relation passing through the median values of each $\sin\theta$ bin of this $\ln(\nu)$ bin is shown in the $R-Z$ plane. In most cases, the best fit ellipes describe the iso-density surface well; some outliers appear at radii where the halo K giant sample is sparser.

\begin{figure}
   \includegraphics[width=\columnwidth]{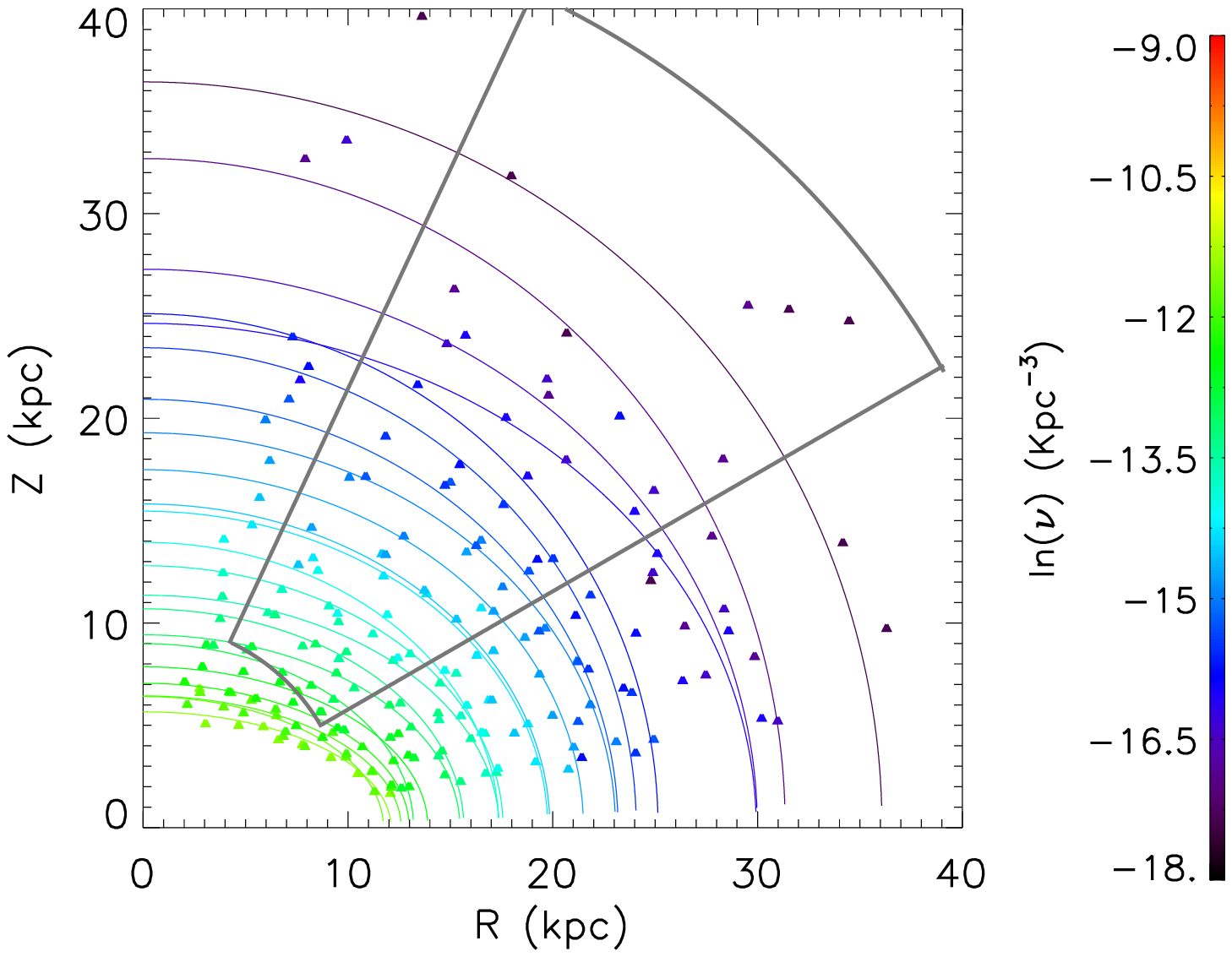}
   \caption{Isodensity contours.  This figure shows all of the best-fit ellipses for each $\ln(\nu)$ (one of which is shown in the right panel of Figure~\ref{RZ_m13.5}). The triangles are the median values similar to those described in Figure~\ref{rsintheta}, but for all iso-density contours.
The colors encode the value of $\ln(\nu)$. The grey wedge indicates the volume explored by SDSS BHB stars of Deason et al. (2011), as comparison. The radial range of SDSS BHB is about 10 to 45 kpc. The latitude range is about $|b|>30^\circ$ (including the substructures) and $30^\circ<|b|<65^\circ$ (excluding the substructures).} 
   \label{RZ}
   \end{figure}

Figure~\ref{RZ} shows the median values and outline of the elliptical relationship $r=\sqrt{R^2+(Z/q)^2}$ for each $\ln(\nu)$ bin, which shows that the ellipses are more oblate at small distances to the Galactic centre and become more spherical at larger distances.

\begin{figure}
   \includegraphics[width=\columnwidth]{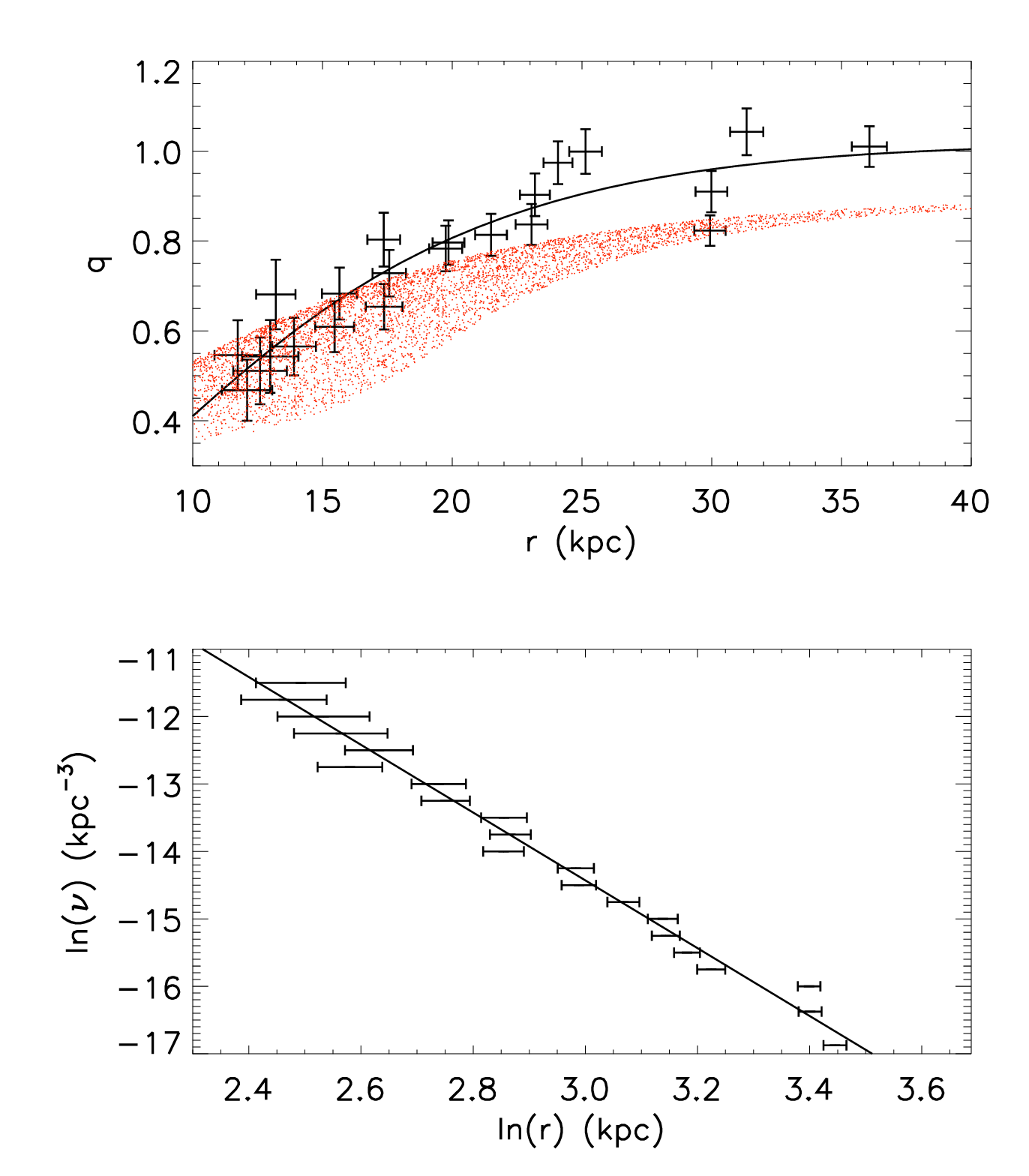}
   \caption{The relationship between the shape parameters $q$ and $r$ at different densities.  The upper panel shows the $q$ distribution as function of $r$.  The black line shows a rational polynomial fit to the data.  The red points show the results from Xue et al. 2015. The lower panel shows the number density as a function of $\ln(r)$. The black line is a power law fit to the data.}
   \label{aa_lnnu}
   \end{figure}

The relationship between $r$ and $q$, as a function of $\ln(\nu)$, is shown in Figure~\ref{aa_lnnu}. The upper panel shows the relationship between $r$ and $q$, and the lower panel shows the relationship between $ln(r)$ and $\ln(\nu)$. The error bars indicate the fitting errors derived from MCMC. We empirically fit the $r, q$ relation with the rational polynomial (black line): $q=\frac{p_1r^2+p_2r}{r^2+p_3r+p_4}$, with $p_1$=0.887, $p_2$=5.15, $p_3$=-7.418, and $p_4$=315.79.
The red points show results from X15, for comparison.
X15 defines $q$ as function of $r_{GC}$ (see Equation 5 of X15). We calculate $r$ and $q$ for each individual K giant of our sample with X15's function and parameters and plot it in red on our figure.
The value of $r$ is different for stars with same $R_{GC}$ but different latitude, so the relationship between $r$ and $q$ is a band in this diagram. Our fit is steeper than X15's result within 25 kpc. Beyond 25 kpc, we find a more spherical halo with $q$ larger than 0.9. 
We didn't fit isodensity contours beyond $r>35$ kpc, because our sample is incomplete at those distances.

The lower panel of Figure~\ref{aa_lnnu} shows the relationship between $\ln(\nu)$ and $\ln(r)$. We fit the relation with a power law. It can be well fitted with a single power law, with best fit power index of $n=-5.03^{+0.64}_{-0.64}$. We also tried to fit the $\ln(\nu)$ and $\ln(r)$ relation with broken power law, but the break radius was larger than 55 kpc, which is beyond the sampling range of our data.

The error bars in the upper and lower panels of Figure~\ref{aa_lnnu} indicate the fitting errors. 
There are other errors which are not included in the fitting errors: 1) The selection correction is a key point of this work. In the worst case, the error of $\ln(\nu)$ can be 0.5-1 for a single plate which introduces a similar fluctuation in the isodensity surface. 2) The random and systematic error in determining the distance is a source of error in the selection correction. 3) Unresolved substructure will introduce bias into the isodensity surface; substructure which does not belong to ``the smooth component'' will make the photometric catalogue denser, and that will in turn cause the number density function to be larger as well. 4) The remaining metal poor disc stars may weakly affect the density at low latitude. We tried to fit the median of the iso-density surface along $\sin\theta$ to reduce the effect of fluctuations in the isodensity surface and of substructure. In the future we will take steps to update of the selection correction to include errors in the estimation of distance.

\subsection{Residuals and substructure}

\begin{figure}
   \includegraphics[width=\columnwidth]{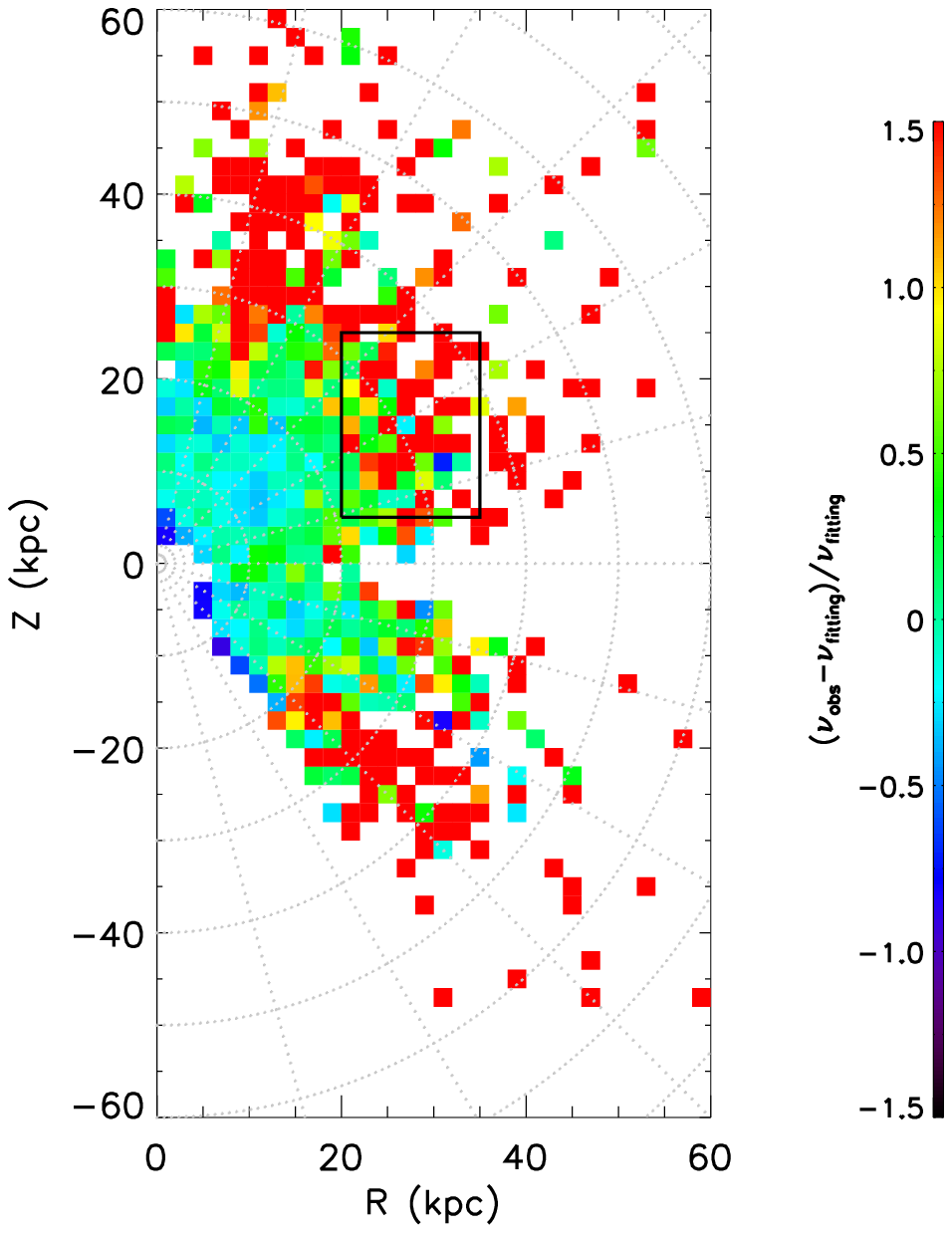}
   \caption{Residual map of the number density, ($\nu-\nu_{fitting})/\nu_{fitting}$, shown in $R$ vs. $Z$ plane.  The most prominent overdensities at the top and bottom of this figure could be due to the Sagittarius dwarf tidal stream in the north and south Galactic hemispheres, respectively. The rectangle indicates the position of overdensity of $(R,Z)=(30,15)$ kpc.}
   \label{residualRZ}
   \end{figure}

The residual map in Figure~\ref{residualRZ} is constructed by calculating residual, $(\nu-\nu_{fitting})/\nu_{fitting}$, for each star and then finding the median residual in each $R-Z$ bin. The bin size is still $2\times2$ kpc as in Figures 4. $\nu_{fitting}$ is obtained from the power law density profile $\nu_{fitting}=(1/r)^n$. The value of $r$ and $q$ for each star is obtained by solving the set of equations: $r=\sqrt{R^2+(Z/q)^2}$ and $q=\frac{p_1r^2+p_2r}{r^2+p_3r+p_4}$. The residual map within 10 kpc can be ignored, because there is no estimate of $q$ and $r$ within 10 kpc, and the extrapolation of $q$ in this range is not reliable. Although we show the residual map up to 60 kpc, we should keep in mind that the map is only complete to 35 kpc for stars with $-4<M_r<-0.5$. Also, $q$ is extrapolated as a constant beyond $r=35$ kpc. We assume that the power law index is unchanged for $r > 35$ kpc.

Figure~\ref{residualRZ} shows that the number density distribution of K giants is well fit within 25 kpc. The median of distribution of residuals is less than 20\% in the region where $10<r_{GC}<25$ kpc. Beyond $r_{GC}=25$ kpc , the number density residual shows some prominent features. The significant feature in the location of $Z>30$ kpc and $R<25$ kpc may be associated with the leading arm of the Sagittarius tidal stream in the northern Galactic hemisphere. The scattered overdensity in the southern sky may associated with the trailing arm of Sgr. We show this in detail in section 3.4 and Figure~\ref{Sgr_candidate}. In addition to the known features, we detect one prominent overdensity around $(R,Z)=(30,15)$ kpc.
We label the overdensity with black rectangle in Figure~\ref{residualRZ}.

\subsection{The overdensities related to the Sagittarius stream}

\begin{figure}
   \includegraphics[width=\columnwidth]{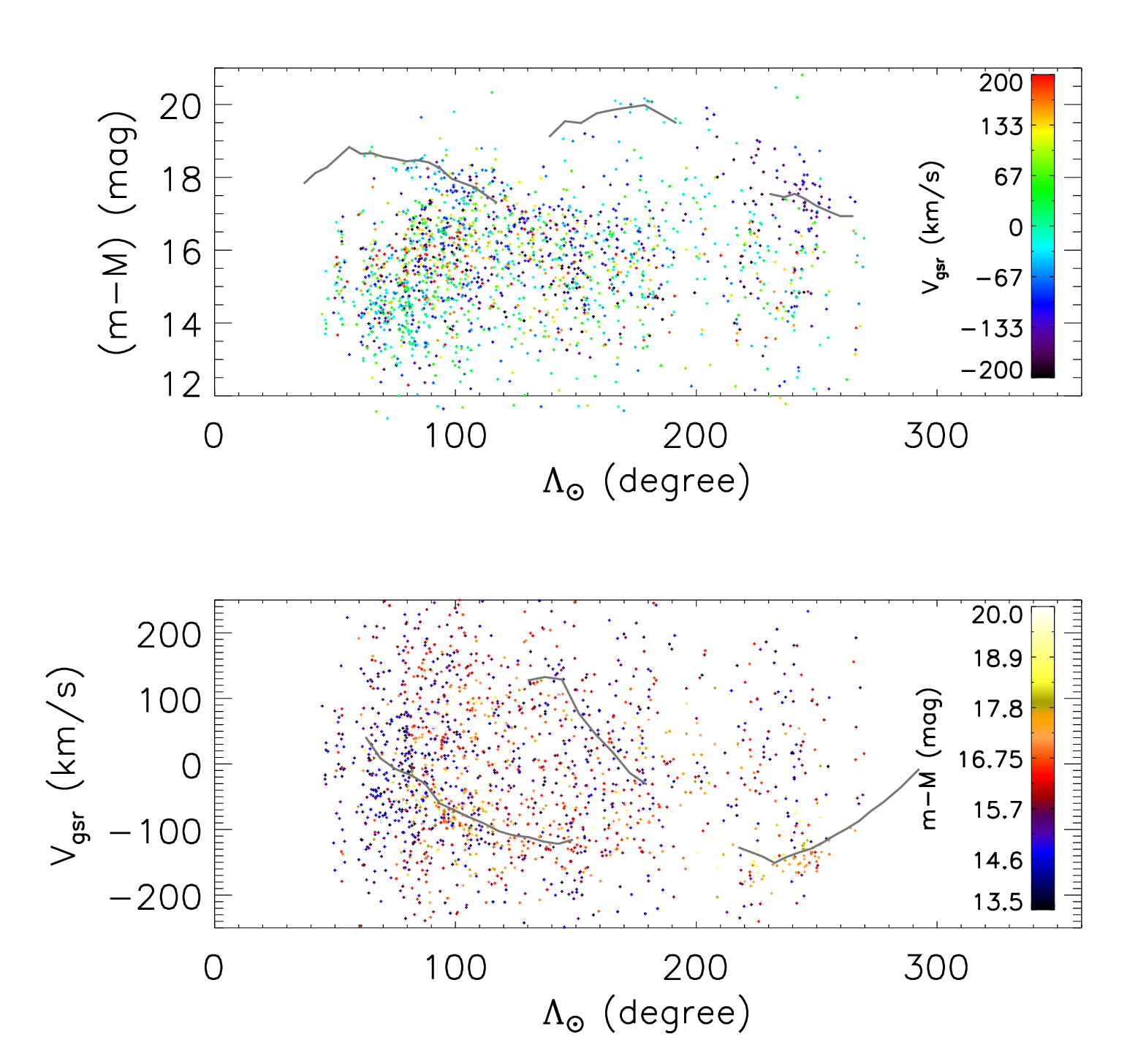}
   \caption{The distance moduli (upper panel) and $V_{gsr}$ (lower panel) for LAMOST K giants with $-12^\circ<B_{\odot}<12^\circ$ as a function of $\Lambda_{\odot}$.  ($\Lambda_\odot$, $B_\odot$) are Sgr stream coordinates defined in \citet{Belokurov2014}. The upper panel shows distance modulus vs. $\Lambda_\odot$, with selected K giants  coded by $V_{gsr}$.  The lower panel shows $V_{gsr}$ vs. $\Lambda_\odot$ with stars colour coded by distance modulus. The grey curves show the heliocentric distance modulus and Galactocentric velocities of Sgr leading and South Sgr trailing arms indicated in Tables 1-5 of \citet{Belokurov2014}. }
   \label{lsgrvsdistandVgsr}
   \end{figure}

\begin{figure}
   \includegraphics[width=\columnwidth]{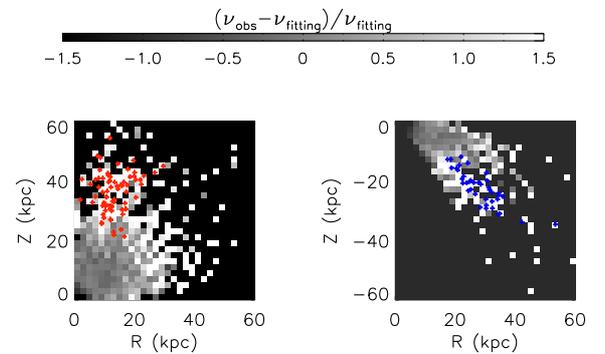}
   \caption{Expected location of Sgr dwarf tidal debris.  Left panel: the red points are candidate stars leading tidal tail stars from the Sgr dwarf galaxy, selected from our LAMOST K giant sample. The background is part of the number density residual map. Right panel: the blue points are candidates stars from the trailing arm of Sgr, selected from our LAMOST K giant sample.}
   \label{Sgr_candidate}
   \end{figure}

\begin{figure}
   \includegraphics[width=\columnwidth]{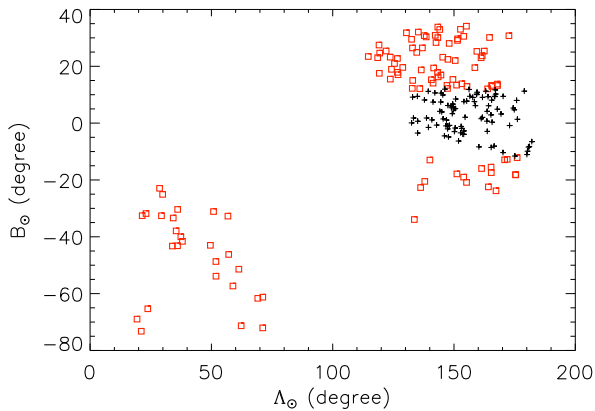}
   \caption{ The distribution of candidates the overdensity at $(R,Z)=(30,15)$ kpc in the Sagittarius tidal stream coordinate system. The black plus signs represent stars within the range $-12^\circ<B_\odot<12^\circ$. The red points are stars beyond this range.}
   \label{lb_RZ30_15}
   \end{figure}

\begin{figure}
   \includegraphics[width=\columnwidth]{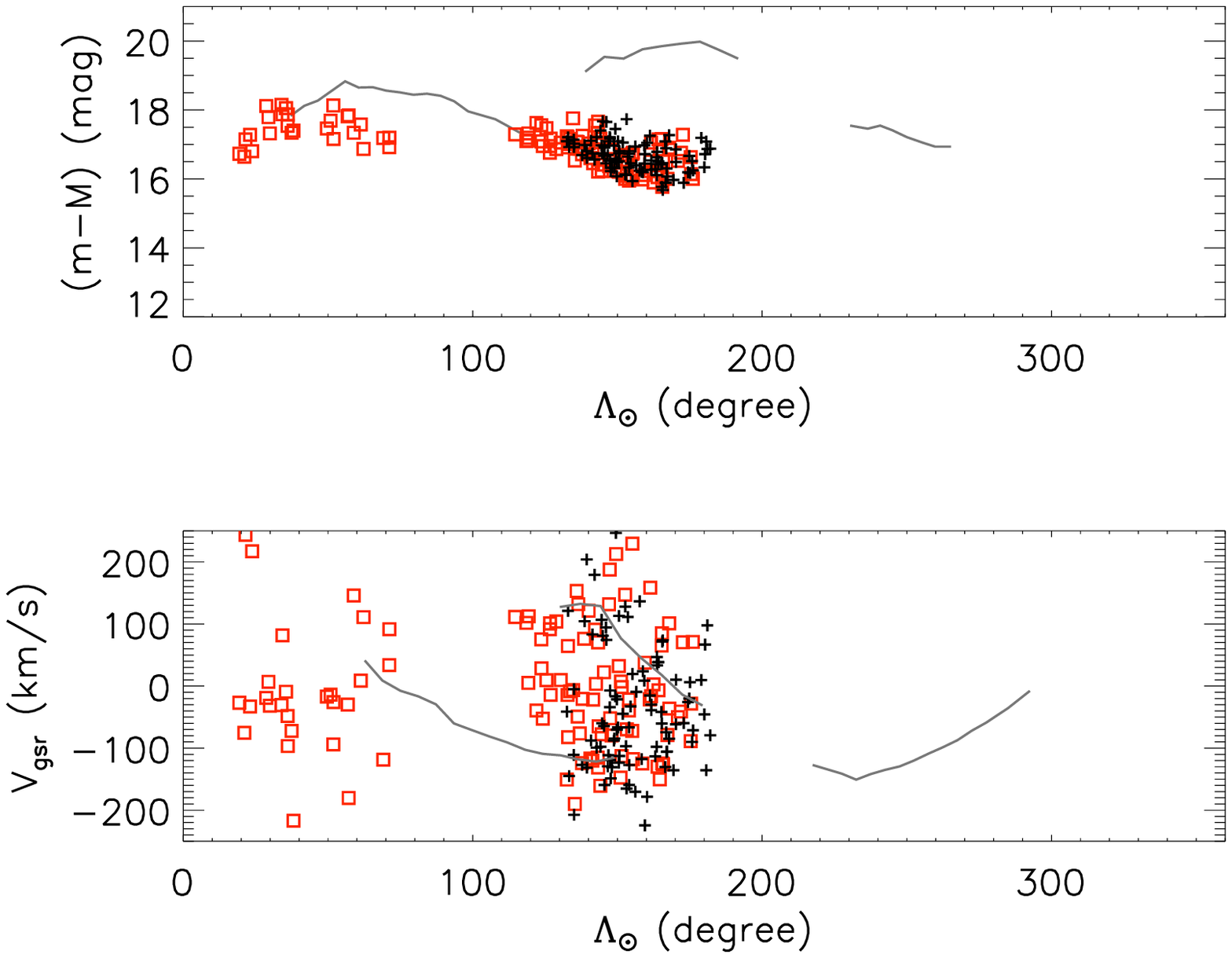}
   \caption{The candidate members of overdensity $(R,Z)=(30,15)$ kpc, plotted in the frame of Figure~\ref{lsgrvsdistandVgsr}. The grey curves are defined in the same way as that of Figure~\ref{lsgrvsdistandVgsr}. The black plus signs represent stars within the range of $-12^\circ<B_\odot<12^\circ$.}
   \label{lsgrvsdistandVgsr_RZ30_15}
   \end{figure}

The prominent overdensities in north and south Galactic hemisphere shown in the residual map may be associated with the leading and trailing arms of Sgr. In this section, we validate the correlation between the identified overdensities in the residual map and the Sgr tidal stream by selecting Sgr candidates from our K giant samples using criteria from \citet{Belokurov2014}, and then overplotting the Sgr candidates on the residual map.

Figure~\ref{lsgrvsdistandVgsr} shows the LAMOST K giants within $\pm12^\circ$ of the equator of the Sgr coordiante system (within $\pm12^\circ$ of the Sgr dwarf orbital plane).
 The figure shows the distribution of $V_{gsr}$ \footnote{$V_{gsr}=rv+9.58*\cos(b)*\cos(l)+(220+10.52)*\cos(b)*\sin(l)+7.01*\sin(b)$ (Xu et al. 2015)} and distance modulus as a function of angle along the tidal stream respectively.
 The coordinate of ($\Lambda_\odot$, $B_\odot$) is a heliocentric coordinate system whose equator is aligned with the Sgr trailing tail. The longitude $\Lambda_\odot$ increases in the direction of Sgr motion, and the latitude $B_\odot$ points to the North Galactic pole, which is the same as defined in \citet{Belokurov2014}.

The paths of the Sgr leading and trailing arms along the distance modulus, as described in \citet{Belokurov2014}, are shown in the upper panel of Figure~\ref{lsgrvsdistandVgsr} as grey curves. The north-leading-arm path starts from ($\Lambda_\odot$, (m-M))=$(37.1^\circ,17.83)$ and ends at $(117.2^\circ,17.3)$. Compared with the leading arm path of \citet{Belokurov2014}, the leading arm traced by LAMOST K giants with $V_{gsr} \sim  -100$ kms$^{-1}$ (blue points) extends to ($\Lambda_\odot$, (m-M))=$(160^\circ,16)$, which is consistent with that traced by the stream's sub-giant branch (SGB) stars reported by \citet{Belokurov2006}. The path of the Sagittarius trailing arm in the north, defined by \citet{Belokurov2014}, starts from ($\Lambda_\odot$, (m-M))=$(138.9^\circ,19.11)$ and ends at $(191.7^\circ,19.49)$, the distance of trailing arm in the north is further than 60 kpc. The path of Sgr trailing arm in the south, as traced by \citet{Belokurov2014} at $220^\circ<\Lambda_\odot<300^\circ$, is observed in LAMOST K giants in the region $220^\circ<\Lambda_\odot<250^\circ$ and $16.5<(m-M)<18.5$.  This identification is confirmed by the obvious clump at ($\Lambda_\odot,V_{gsr}$)=$(245^\circ,-150$ kms$^{-1})$ in lower panel of Figure~\ref{lsgrvsdistandVgsr}.

Note that the Sgr trailing arm in the north is farther than 60 kpc, which is beyond our detection range, so it won't be considered seriously in this work.

We select Sgr northern leading arm and southern trailing arm candidates along the path of $V_{gsr}$ and distance modulus as a function of $\Lambda_\odot$ as defined by \citet{Belokurov2014}. The northern leading arm candidates are selected using $V_{gsr}$ within 50 kms$^{-1}$ and $(m-M)$ within 1 mag of the Sgr leading arm paths in Figure~\ref{lsgrvsdistandVgsr}. The southern trailing arm candidates are also selected along the corresponding path depicted in Figure~\ref{lsgrvsdistandVgsr}. Because the clump of LAMOST K giants at ($\Lambda_\odot,V_{gsr}$)=$(245^\circ,-150$ kms$^{-1})$ is slightly shifted from the grey curve of the Sgr southern trailing arm in the lower panel of Figure~\ref{lsgrvsdistandVgsr}, the distance between $V_{gsr}$ of LAMOST K giants and the path centre is required to be greater than -70 kms$^{-1}$ and smaller than 30 kms$^{-1}$ within the longitude range $230^\circ<\Lambda_\odot<255^\circ$.

The selected Sgr northern leading arm candidates are overplotted in the left panel of Figure~\ref{Sgr_candidate}, while the southern trailing arm candidates are labelled in the right panel. The background is part of residual map of Figure~\ref{residualRZ}. The Sgr candidates lie exactly in the regions where the observational number density is higher than the best fit model.

We also tried to select candidates associated with the overdensity $(R,Z)=(30,15)$ kpc, within the distance range $20<R<35$ kpc, $5<Z<25$ kpc, as depicted by the rectangle in Figure~\ref{residualRZ}. The candidates are selected with $(\nu-\nu_{fitting})/\nu_{fitting}>0.5$. To determine whether the overdensity is associated with Sgr, we plot the candidates in the Sgr coordinate system in Figure~\ref{lb_RZ30_15}. It is seen that there are two separated sub-groups of candidates. The first one, with $\Lambda_\odot<80^\circ$ and $B_\odot<-20^\circ$, is far from the equator of the Sgr stream coordinate system. The second one, with $110^\circ<\Lambda_\odot<180^\circ$ and $-20^\circ<B_\odot<40^\circ$, passes through the Sgr equator.
Figure~\ref{lsgrvsdistandVgsr_RZ30_15} shows the distribution of distance modulus and $V_{gsr}$ for candidates associated with the overdensity at $(R,Z)=(30,15)$ kpc as a function of distance along Sgr longitude. In the upper panel of Figure~\ref{lsgrvsdistandVgsr_RZ30_15}, the overdensity candidates are distributed along the extension of the line tracing the Sgr northern leading arm. In the lower panel of Figure~\ref{lsgrvsdistandVgsr_RZ30_15}, the $V_{gsr}$ of part of the candidates overlap the grey curve of Sgr northern leading arm. However, many other the candidates are surprisingly distributed closer to the grey curve of north-trialing-arm. The $V_{gsr}$ distribution of the overdensity at $(R,Z)=(30,15)$ kpc is clearer in Figure~\ref{lsgrvsdistandVgsr_RZ30_15_mpoor}, showing the metal poor subsample, in APPENDIX A. The distance to the overdensity at $(R,Z)=(30,15)$ kpc is about 25 kpc, while the northern trialing arm is further than 60 kpc; the distances are not consistent with each other. So the overdensity cannot be a piece of northern trialing arm.
It might be part of Sgr in other wrap, or, it might be a new subsucture. Because the origin of the overdensity is beyond the scope of this paper, we will pursue this subject in  the further work.

\section{Discussion and Conclusion}

Carollo et al.(2007) found that the stellar halo can be separated into the inner and outer halos based on density profile, kinematic, and metallicity. The boundary between the inner and outer halos is about 15 kpc. The inner halo is more oblate than the outer halo. In addition, the inner and outer halos may have different formation histories. The accretion of the inner halo is thought to have happened 10 Gyr ago, and part of inner halo may have formed {\it in situ}. The accretion of the outer halo happened much more recently than that of inner halo; most of the outer halo is thought to be composed of accretion debris (Abadi et al. 2006, Bland-Hawthorn et al. 2016).

In later studies, the density profile of the halo has been mapped with different tracers (Deason et al.2011, Sesar et al. 2010). They found the inner halo has a shallower density profile than outer halo. The simulation of Deason et al. (2013) indicates that if there is a break in the power law index, then the strength of the break depends on the accretion time and the size of the accretion debris; the location of the break depends on the average apocentre of the accreted halo component.

We presnt a direct method to detect the structure of the Galactic halo.
The method relies on relatively complete sky coverage (including areas of both high Galactic latitude and low Galactic latitude) and a large sample of spectroscopic tracers.
In our work, we started from the defination of density distribution and fit the contours of the iso-density surface. We did not predetermine whether the flattening would change with Galactic radius, whether there is break on the power law index, or where is the possible break might occur.
We determined the density profile and flattening by fitting the iso-density surface in the $(r_{GC},\sin\theta)$ plane. 
We did not eliminate whole areas of the sky occupied by stellar streams; we fit medians of the iso-density surface to avoid the influence of outliers which could be related to substructures.
Because we used the median value of ($r_{GC}$,$\sin\theta$) when we fit the iso-density surface in the $r_{GC}-\sin\theta$ plane, the affect of the tidal streams was reduced, though not entirely removed.

We obtained the flattening and power law index independently, breaking the degeneracy between flattening and power law index. We find that the flattening $q$ changes with Galactic radius, which is consistent with result of the SPL model of X15; $q$ changes rapidly from $r=15$ kpc to 25 kpc. We find no evidence of a break of power law between 15 kpc and 35 kpc.
Our data does not extend as far as X15's SDSS K giants, but our sample is much larger. Because of the large sample and the fact that we adopted a non-parametric method, the constraint on the flattening is more independent and reliable in the distance range ($r_{GC}<$ 35 kpc).

To show the difference between our result and previous results more clearly, we compare them further. From table 1, some of previous works support SPL (Robin, et al. 2008, Juric et al. 2008, Bell et al. 2008). Meanwhile, some of others support BPL (Watkins et al. 2009, Sesar et al. 2011, Deason et al. 2011, Pila-Diez et al. 2015). The broken position changes from 19 kpc to 28 kpc, but the power law index of inner and outer halo are quite similar with each other. The density can be fitted either by SPL with varying q or BPL in the work of X15 and Das et al. (2016). Other recent works support triaxial halo model (Pila-Diez et al. 2015, Iorio et al. 2017). 

\begin{figure}
   \includegraphics[width=\columnwidth]{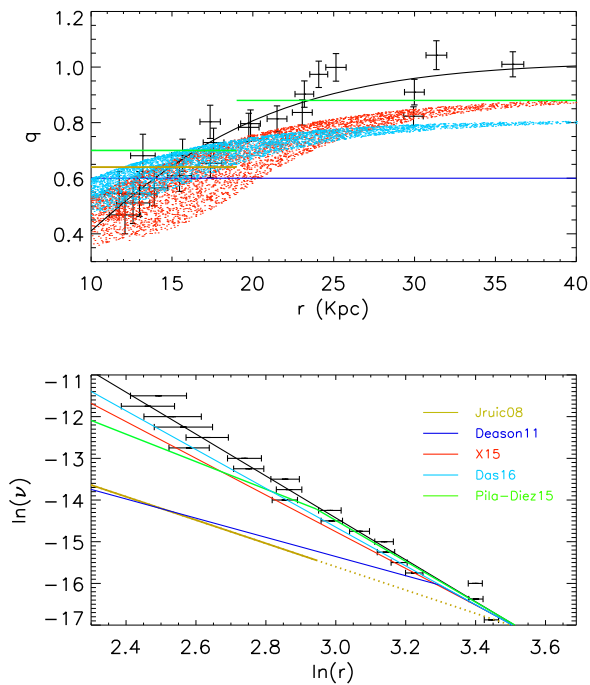}
   \caption{Same as Figure~\ref{aa_lnnu}. The flattening and power law index of previous works are overplotted. The black points and black fitting curve and line represent the result of this work. The red points in the upper panel and red line in the lower panel shows the result of SPL with varying q from X15, while the light bule lines represent the result of SPL with varying q from Das et al.(2016). Note that the functions of $q(r_{GC})$ are different in X15 and Das et al. (2016). The blue lines represent the result from Deason et al. (2011). The brown lines represent Juric et al. (2008)'s result, the brown dots shows the extention line of Juric et al. (2008)'s result. The green lines represent result of BPL with segmental q from Pila-Diez et al.(2015). In the lower panel, the zero point of $ln(\nu)$ distribution of previous works is arbitrary.} 
   \label{compare_index}
   \end{figure}

The typical results of previous works are compared with our result in Figure~\ref{compare_index}. Because our analysis is under the assumption that the halo is axisymmetric. It's hard to compare with results of fitting by triaxial halo models. From the comparison in figure~\ref{compare_index}, our result is much steeper than the other results fitted with either BPL (Deason et al. 2011, blue lines) and SPL (Juric et al. 2008, brown lines) within the range of 25 kpc. The density profile modelled by BPL with segmental flattening (Pila-Diez et al. 2015, green lines) is steeper than Deason et al. (2011) and Juric et al. (2008) but flatter than our result. The result from X15 (red points and lines) and Das et al. (2016, light blue points and lines) fitting by SPL with varying q are similar to this work. 

Compared to previous works, we adopted different tracers and apply different approach, which may lead to systematic difference in the slope of the power-law model. However, the assumption of a constant q with value smaller than 1 and the selection effect of excluding the low Galactic halo stars may be the main reasons that the slope is much flatter in many of the previous works.

Figure~\ref{densityprofile} and the discussion in section 2.3 provide two exterme cases of the stellar density profile, one is along $ln(R)$ and the other along $ln(Z)$. It shows that when q is variable the two profiles have different slopes. If an inproper assumption that q is a constant is made, then the observed stellar density profile should be somewhere in between the two extreme profiles shown in Figure~\ref{densityprofile}. Consequently, the best fit slope of the data should be substantially flatter than the slope derived from the profile along $ln(R)$ (the black symbols). Note that the parameter r in our model represent for the length of the semi-major axis, which is equivalent to R when q<1. Therefore, the derived slope with the assumption of a constant q smaller than 1 would lead to a flatter slope of the power-law than our result. Moreover, almost all previous works excluded the low Galactic latitude halo stars to aviod disc contamination. This leads to a systematics bias toward the density profile along $ln(Z)$, which is flatter than that along $ln(R)$ (see Figure~\ref{densityprofile}). Thus, again, the selection effects prefers to a flatter power-law. Combining the two situations together, it is not surprise that many previous works using constant q derived much flatter slope, whileas X15 and Das et al.(2016), who adopted a variable q, obtain steeper slope similar to this work.

Our work measure the iso-density of the halo directly and model independent. The most crucial ingredients relies on the selection correction.
We plan to revisit this technique in the future, with an updated selection correction and combining Pan-STARRS and UCAC4 photometry, so that the completeness will be extended to a fainter magnitude limit. Also, in LAMOST DR4, the $\alpha$ abundance will be provided so that we can calculate better distance estimates for the giant stars. The proposed improvements will allow us to better study the structure of the halo, particularly for $r_{GC}>35$ kpc.

\section*{Acknowledgements}
This work is supported
by the Strategic Priority Research Program ``The Emergence of Cosmological Structures'' of the Chinese
Academy of Sciences, Grant No. XDB09000000 and the National Key Basic Research Program of China
2014CB845700.
CL acknowledges the National Natural Science Foundation of China grant No. 11373032 and 11333003. HJN acknowledges support from the US NSF grant AST 16-15688. Guoshoujing Telescope (the Large Sky Area Multi-Object Fiber Spectroscopic Telescope LAMOST) is a National Major Scientific Project built by the Chinese Academy of Sciences. Funding for the project has been provided by the National Development and Reform Commission. LAMOST is operated and managed by the National Astronomical Observatories, Chinese Academy of Sciences. 
We appreciate the anonymous referee for his/her the very helpful comments and suggestions.


\appendix

\section{Checking our result with metal-poor subsample}

In the main body of this paper,
we use LAMOST K giants with [Fe/H]$<-1$ to trace the density profile of the halo. This criterion, and the exclusion of the  $\sin\theta=0.1$ bin when we fit the density profile of halo in Figure ~\ref{rsintheta}, removes most of the disc stars. But our sample might still include contamination from a metal-poor thick disc component (X15). Although we don't expect it to inflence our result much, we check our result with more metal poor sample which is more likely a pure halo sample in this Appendix.

\begin{figure}
   \includegraphics[width=\columnwidth]{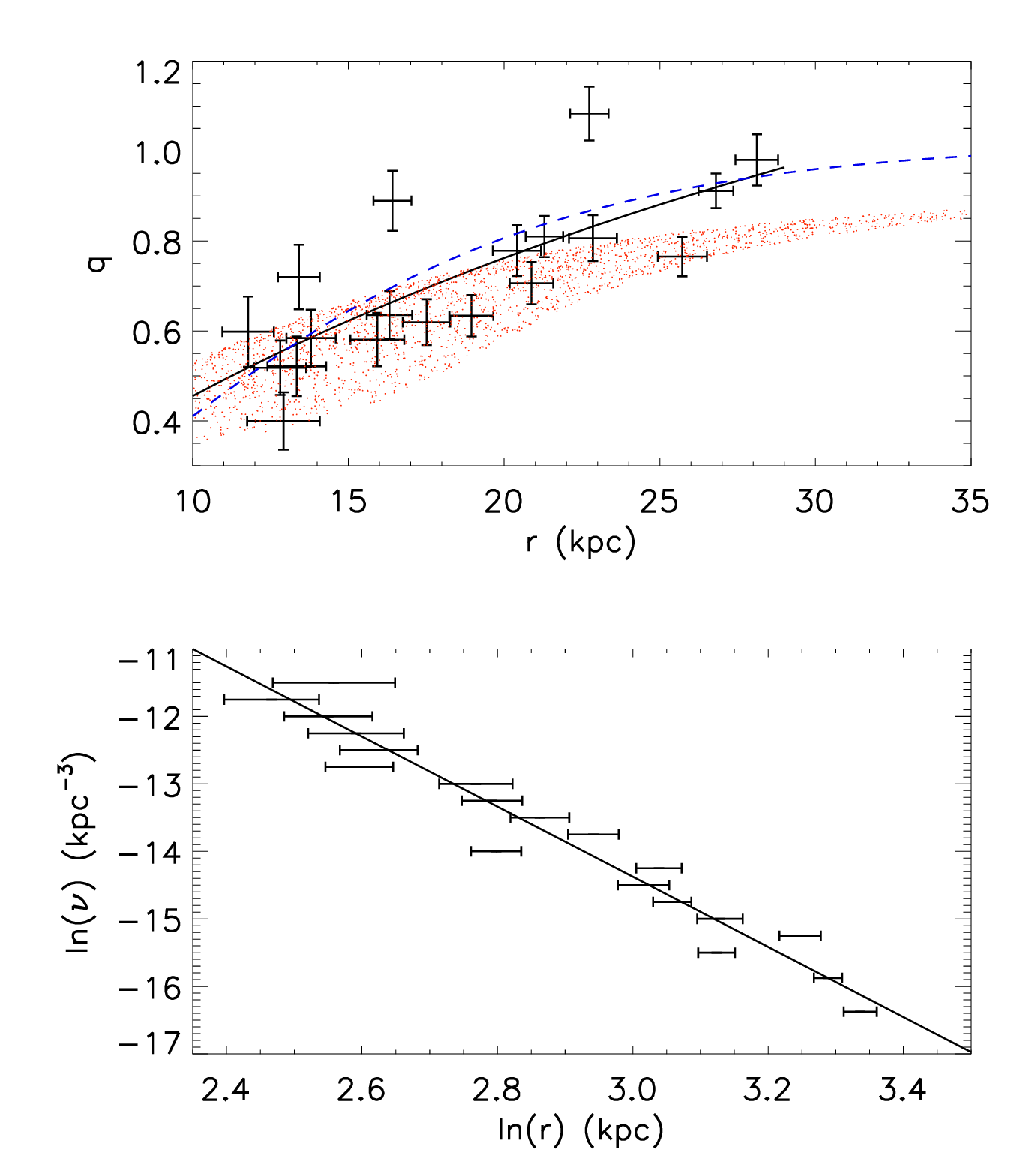}
   \caption{Same as Figure~\ref{aa_lnnu}, but with [Fe/H]$<-1.5$. The upper panel shows the $q$ variation with $r$. 
The black curve is the fitting result. The blue dashed curve, same as the black curve in Figure ~\ref{aa_lnnu},  is the fit result for the total sample as a comparison. The red points represent result of Xue(2015). The lower panel shows the number density of K giants as a function of $\ln(r)$. }
   \label{aa_lnnu_mpoor}
   \end{figure}

We limit the subsample of K giants to [Fe/H]$<-1.5$ in this analysis; the lower metallicity cut results in a sample of 2484 K giants, which is about half of the previously analyzed sample. We fit each iso-density surface for the free parameters $r$ and $q$ with the same method used in the larger sample. Figure ~\ref{aa_lnnu_mpoor} shows the flattening variation and density profile obtained from the metal-poor subsample. The upper panel shows how the flattening changes with the radius. The axis ratio increases from 15 kpc to 30 kpc, but the uptrend is a little more gentle than that of the larger sample. The flattening q is about 0.62, 0.76, and 0.98 for $r=15, 20,$ and $30$ kpc, respectively.
The power law index $n=-5.19^{0.85}_{0.85}$ is consistent with that of larger sample.

Because the subsample of [Fe/H]$<-1.5$ is only half of total sample, it can only trace the isodensity surface within $r<30$ kpc. Futher than that, there are not enough stars in each $\ln(\nu)$ bin. Besides, the scatter in the distribution of axial ratio and power law index observed in Figure~\ref{aa_lnnu_mpoor} is larger than that of Figure~\ref{aa_lnnu}.

\begin{figure}
   \includegraphics[width=\columnwidth]{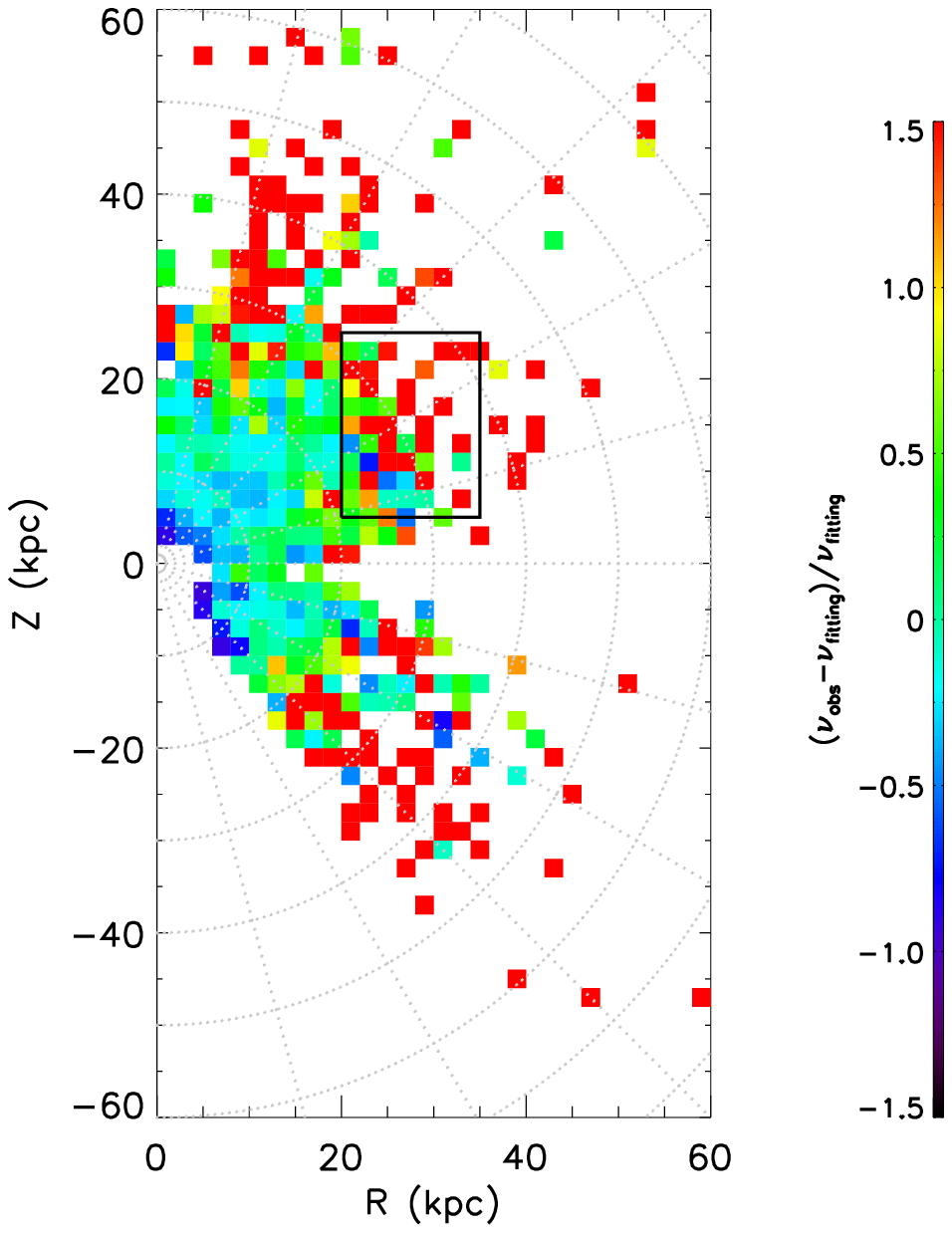}
   \caption{The residual map for stars with [Fe/H]$<-1.5$. The -coded residual of $(\nu-\nu_{fitting})/\nu_{fitting}$ is shown in the $R-Z$ plane. The rectangle shows the position of overdensity $(R,Z)=(30,15)$ kpc.}
   \label{residualRZ_mpoor}
   \end{figure}

We follow the same method used in the larger sample to calculate the residual map of the stellar density.
Figure ~\ref{residualRZ_mpoor} is the residual map for the metal-poor subsample. The results are consistent with that of the larger sample. The median of residual is less than 27\% in the range $r_{GC}<25$ kpc.  Beyond 25 kpc, the overdensities of the Sgr northern leading arm, Sgr southern trialing arm, and the overdensity at $(R,Z)=(30,15)$ kpc are still significant in the map.

\begin{figure}
   \includegraphics[width=\columnwidth]{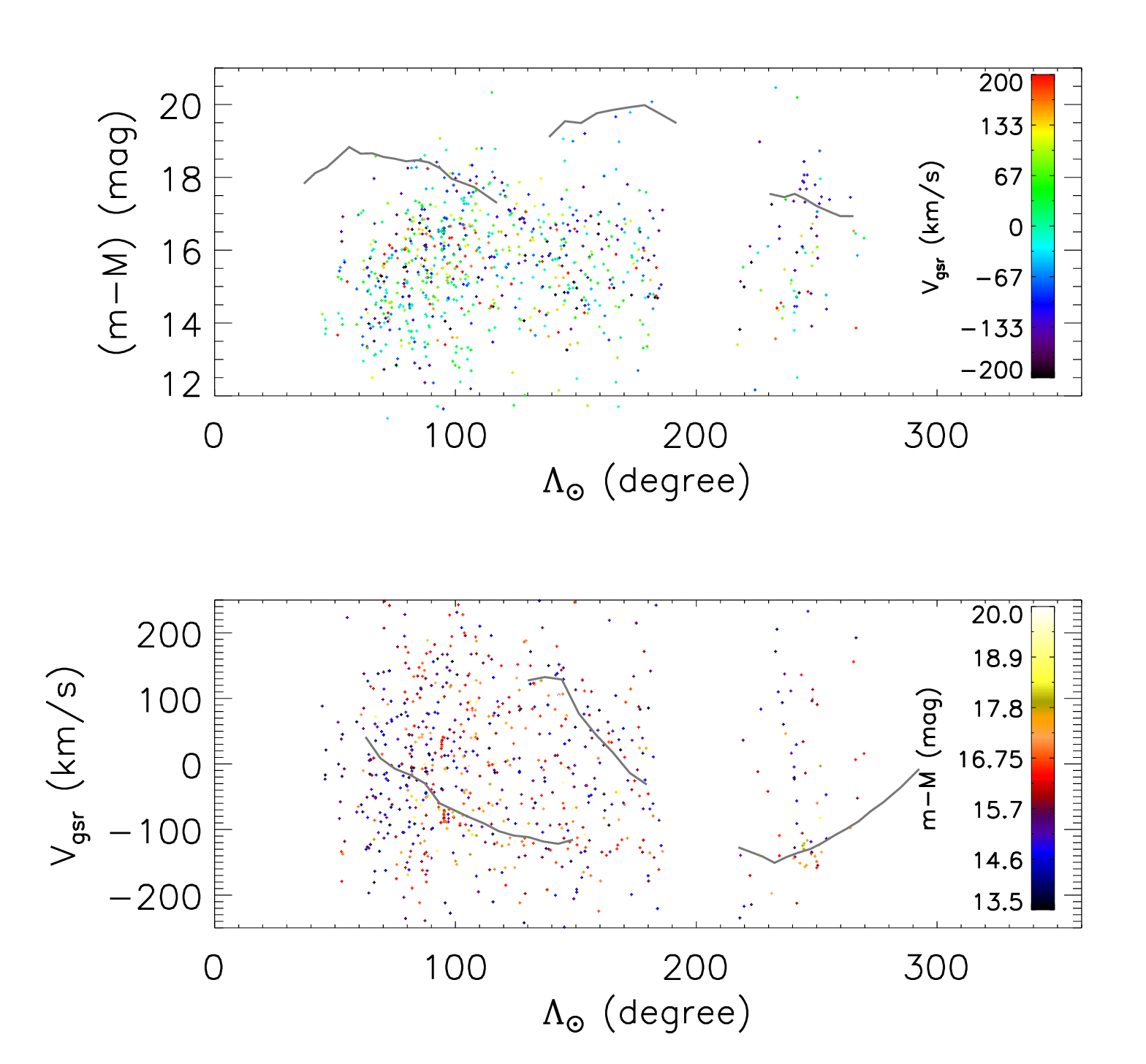}
   \caption{Same as Figure~\ref{lsgrvsdistandVgsr}, but for [Fe/H]$<-1.5$. The points represent K giants with [Fe/H]$<-1.5$ near the Sgr orbital plane, $-12^\circ<B_\odot<12^\circ$. The grey curves represent Sgr leading and trailing arms. }
   \label{lsgrvsdistandVgsr_mpoor}
   \end{figure}

\begin{figure}
   \includegraphics[width=\columnwidth]{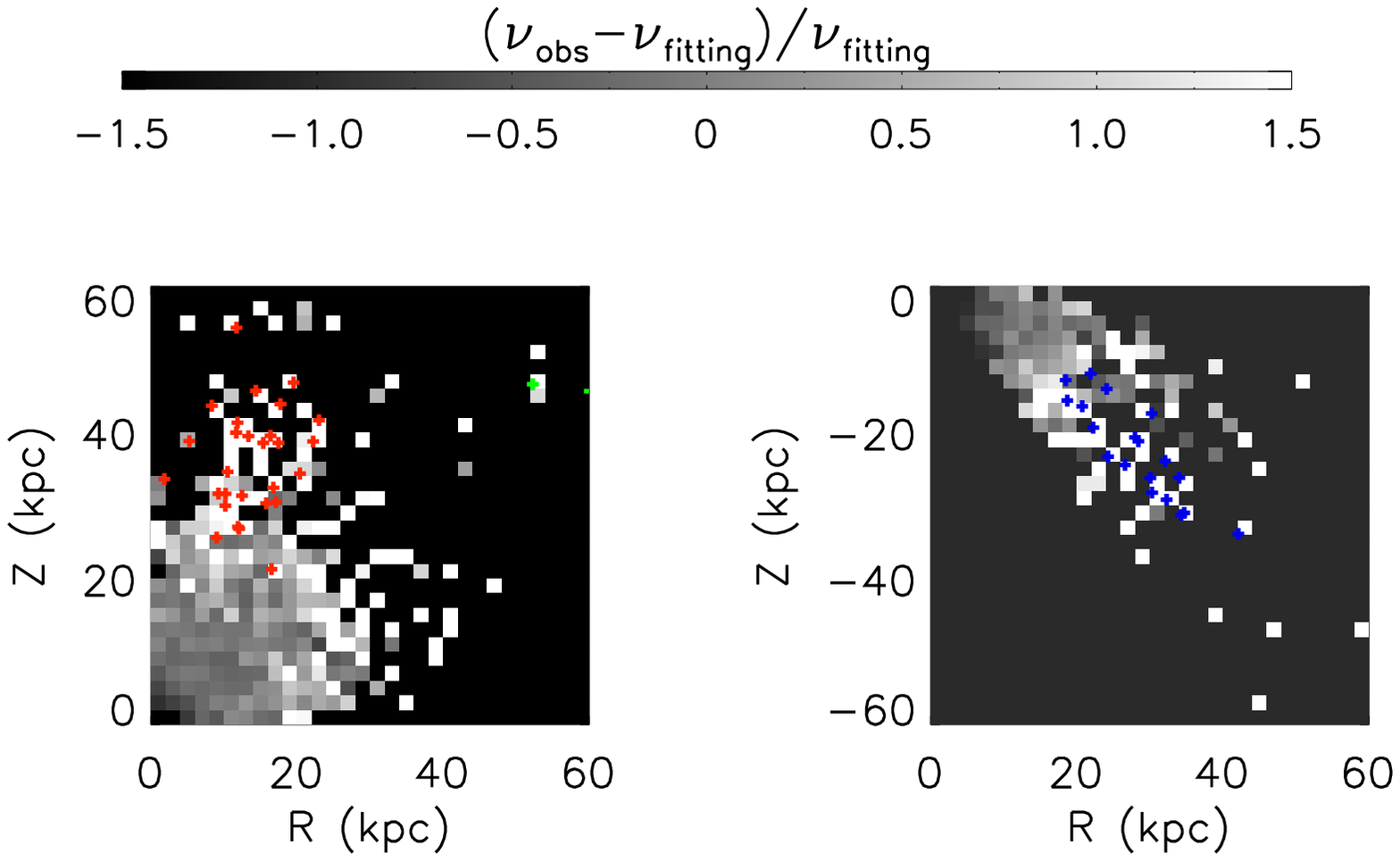}
   \caption{The candidates of Sgr northern leading arm and southern trailing arm in the LAMOST K giant sample with [Fe/H]$<-1.5$, overplotted on the residual map which is part of Figure~\ref{residualRZ_mpoor}. The red points in the left panel show the Sgr northern leading arm candidates. The blue points in the right panel show the Sgr southern trailing arm candidates.}
   \label{Sgr_candidate_mpoor}
   \end{figure}

To study the relationship between the overdensities and the largest substructure, Sagiattarius, we replot Figure~\ref{lsgrvsdistandVgsr} with the metal poor subsample in Figure ~\ref{lsgrvsdistandVgsr_mpoor}.
The patterns are quite similar to those of  Figure ~\ref{lsgrvsdistandVgsr}.
In Figure~\ref{Sgr_candidate_mpoor}, the candidates in the metal-poor subsample also overlap the overdensities in the residual map, which are consistent with those in the larger sample.

\begin{figure}
   \includegraphics[width=\columnwidth]{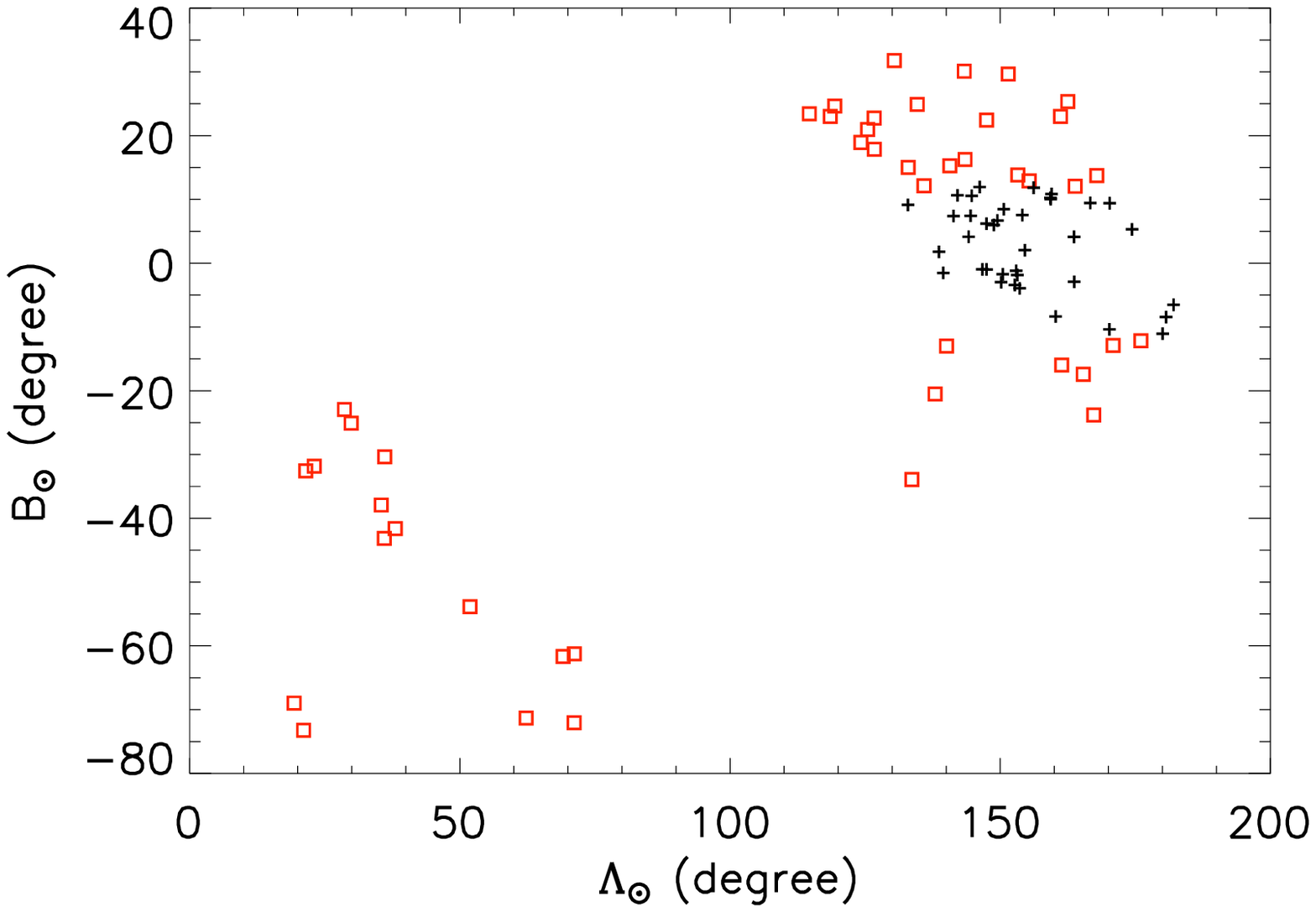}
   \caption{Same as Figure~\ref{lb_RZ30_15}, but with [Fe/H]$<-1.5$. The plus signs represent candidates in the overdensity at (R,Z)=(30,15) kpc. The selection criteria are $20<R<35$ kpc, $5<Z<25$ kpc, $(\nu-\nu_{fitting})/\nu_{fitting}>0.5$. The black plus signs label the stars located within $-12^\circ<B_\odot<12^\circ$; the red points represent stars beyond this range.}
   \label{lb_RZ30_15_mpoor}
   \end{figure}

\begin{figure}
   \includegraphics[width=\columnwidth]{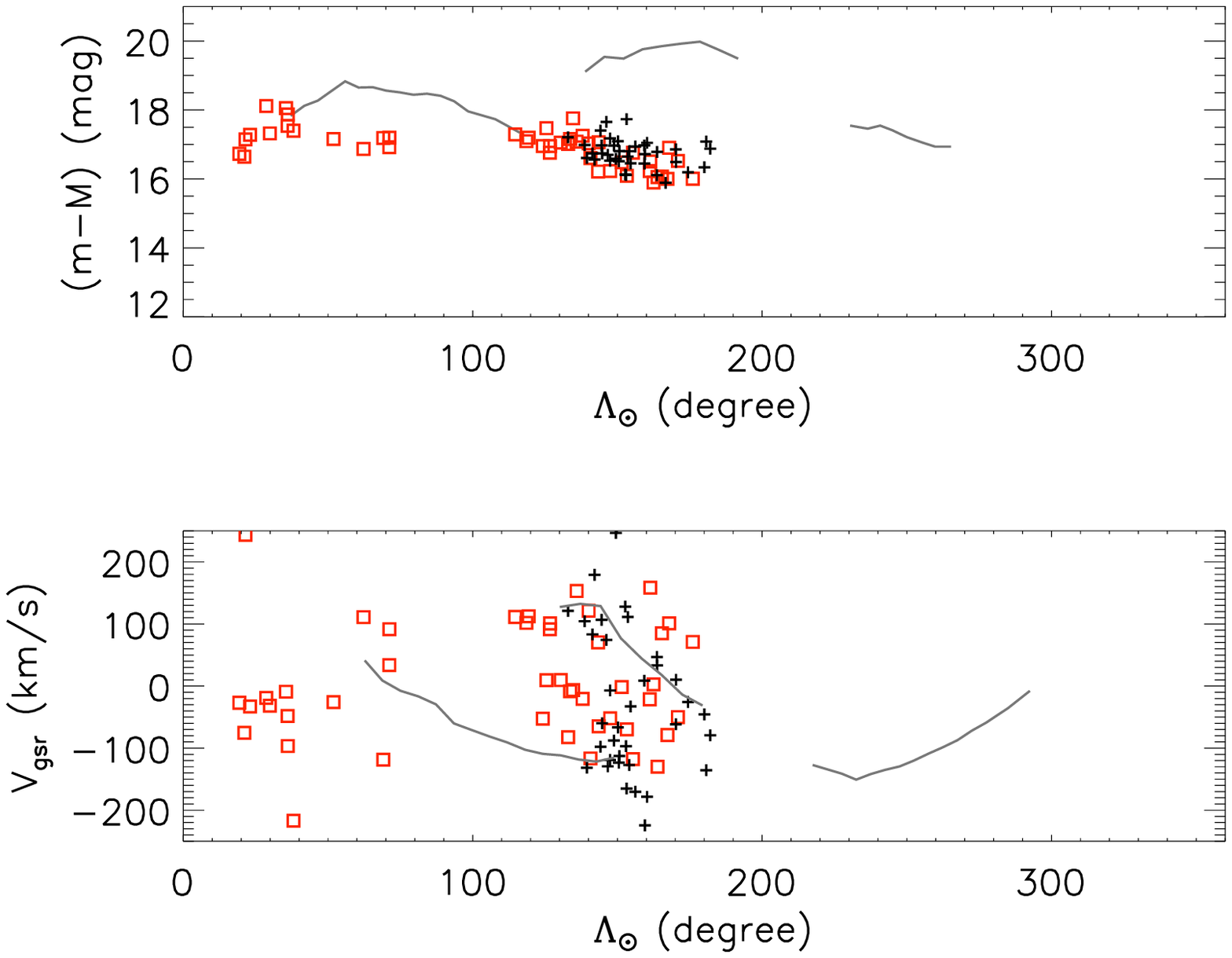}
   \caption{Same as Figure~\ref{lsgrvsdistandVgsr_RZ30_15}, but for [Fe/H]$<-1.5$. The candidates in the overdensity at $(R,Z)=(30,15)$ kpc with [Fe/H]$<-1.5$ are overplotted in the same frame as Figure~\ref{lsgrvsdistandVgsr_RZ30_15}. The red points and black plus signs are similar to those in Figure~\ref{lb_RZ30_15_mpoor}.}
   \label{lsgrvsdistandVgsr_RZ30_15_mpoor}
   \end{figure}

We also select the candidates associated with the overdensity at (R,Z)=(30,15) kpc.
The candidate selection criteria are the same as that of the larger sample and the selected candidates are shown in Figures~\ref{lb_RZ30_15_mpoor} and ~\ref{lsgrvsdistandVgsr_RZ30_15_mpoor}. The metal-poor subsample also shows two groups in Figure~\ref{lb_RZ30_15_mpoor}. The $V_{gsr}$ distribution of metal-poor subsample candidates, in the lower panel of Figure~\ref{lsgrvsdistandVgsr_RZ30_15_mpoor}, shows quite clearly that some of the candidates follow the path of Sgr northern leading arm, and some of the candidates follow the path of Sgr northern trailing arm, in agreement with the findings from the larger sample.

\bsp	
\label{lastpage}
\end{document}